# Bridging LMS and generative AI: dynamic course content integration (DCCI) for enhancing student satisfaction and engagement via the ask ME assistant

Kovan Mzwri[1] · Márta Turcsányi-Szabo[2]

## Abstract

Integration of Large Language Models (LLMs) with Learning Management Systems (LMSs) can enhance task automation and accessibility in education. However, hallucination where LLMs generate inaccurate or misleading information remains a challenge. This study introduces the Dynamic Course Content Integration (DCCI) mechanism, which dynamically retrieves course content from Canvas LMS and structures it within an LLM's context window via prompt engineering, enabling the LLM-powered assistant, Ask ME, to deliver context-aware, curriculum-aligned responses while mitigating hallucinations. A mixed-methods pilot study grounded in Self-Determination Theory (autonomy, competence) and the Technology Acceptance Model (perceived usefulness, ease of use) evaluated DCCI's effectiveness with 120 first-year programming students at Eötvös Loránd University. The course focused on foundational programming patterns in C#, including writing program specifications. We analyzed 14,746 logged interactions and a post-course survey completed by 101 students. User satisfaction was measured via a 5-point Likert scale (turn-level ratings), while the survey assessed usability, engagement, and ethical concerns. Results indicated high satisfaction (mean 4.65/5) and strong recognition of Ask ME's ability to provide timely, contextually relevant answers to administrative and course-related queries. 78.06% agreed that Ask ME's Canvas integration reduced platform switching, improving usability, engagement, comprehension, and topic exploration. Many students reported reduced hesitation to ask questions and increased motivation for self-directed learning, though concerns about over-reliance on AI and reduced student–teacher interaction emerged. This study demonstrates that DCCI enhances LLM reliability, student satisfaction, and engagement in AI-driven educational automation, while highlighting the importance of balancing AI efficiency with pedagogical integrity and human–AI collaboration.

---

Extended author information available on the last page of the article

## Introduction

The integration of artificial intelligence (AI) in education has ushered in a transformative era, offering opportunities to enhance learning experiences, streamline administrative tasks, and promote personalized education (Ahmad et al., 2022; Chen et al., 2020; Slade et al., 2024; Walter, 2024). Among the most promising advancements are Large Language Models (LLMs), which can process and generate human-like text, providing educators and students with dynamic assistance (Bonner et al., 2023; Meyer et al., 2023).

Integrating LLMs with course content presents a promising approach to automate tasks such as education content delivery, dynamic 24/7 student support, grading, assessments, and real time human like feedback content (Alkafaween et al., 2025; Bonner et al., 2023; Morris et al., 2024; Yang et al., 2024). This automation allows educators to concentrate on teaching rather than on repetitive administrative duties, thereby improving overall efficiency and productivity (Rengers et al., 2024). Moreover, by incorporating course-specific information into LLMs and Generative AI (GAI), these models can provide personalized, context-aware guidance that enhances student engagement, supports self-regulated learning, and improves comprehension of complex material (Mittal et al., 2024; Peláez-Sánchez et al., 2024).

Despite these advantages, several challenges must be addressed to effectively implement LLMs in educational settings. A notable concern is the phenomenon of hallucination, whereby LLMs may generate factually inaccurate or irrelevant information (Gravina et al., 2024; Park et al., 2024; Shah et al., 2024). To mitigate this issue, it is essential to establish a robust mechanism that reliably links course content with the LLM, ensuring that generated information is accurate, relevant, and consistent with the curriculum.

To address both the opportunities and limitations of LLMs, this research introduces the Dynamic Course Content Integration (DCCI) mechanism—a novel approach to automating the retrieval and integration of course content from Canvas Learning Management System (LMS) into an LLM's context window. DCCI leverages Canvas's Learning Tools Interoperability (LTI) and Application Programming Interface (API) to dynamically retrieve content from structured course-related resources, such as Canvas pages. The retrieved content is then structured using prompt engineering, combining the relevant context with user queries in the LLM context window to ensure that the LLM receives well-structured, context-aware input. This process enhances the precision, relevance of human-like responses generated by the LLM, effectively reducing hallucination and ensuring alignment with course materials.

The DCCI mechanism is implemented in the Ask ME Assistant Bot, a practical tool developed as part of this study with full integration into Canvas LMS. By leveraging up-to-date course information, the Ask ME Assistant Bot delivers accurate, relevant, and timely responses, addressing critical challenges such as maintaining response accuracy and adapting to weekly curriculum topics and real-time updates. In addition to facilitating access to course content, the bot fosters student engagement and supports self-directed learning, thereby aligning with contemporary educational paradigms.

Guided by Self-Determination Theory (SDT) and the Technology Acceptance Model (TAM), this study addresses:

## Research questions

This study addresses the following research questions:

1. **RQ1**: How does the DCCI mechanism improve the accuracy, relevance and timeliness of LLM-generated responses in an LMS environment?
2. **RQ2**: What is the impact of DCCI-powered Ask ME Assistant on student satisfaction, engagement and perceived educational benefits?
3. **RQ3**: How do students perceive the risk and challenges of AI-driven tools in educational settings?

## Research objectives

This study aims to advance the automation of educational tasks through the integration of LLMs. The primary objectives are:

1. Develop the DCCI mechanism to retrieve and integrate course content and curriculum from the Canvas LMS, facilitating contextually aware responses generated by a Large Language Model (LLM) through the Ask ME Assistant Bot.
2. Assess the effect of DCCI on connecting course content from Canvas LMS to the LLM, and evaluate the accuracy, relevance, and timeliness of responses generated by the Ask ME Assistant Bot within the Canvas LMS environment.
3. Investigate students' perceptions of AI tools (Ask ME) using SDT (autonomy, competence) and TAM (perceived usefulness, ease of use), focusing on satisfaction and engagement, and to examine the perceived benefits and challenges of GAI integration in learning.

## Research hypotheses

1. **H1:** Autonomy support (e.g., 24/7 access to Ask ME, self-directed query resolution) and competence enhancement (e.g., Context-aware responses generated by DCCI) will positively correlate with student satisfaction and reduced hesitation to ask questions or seek academic support (SDT).
2. **H2**: Perceived usefulness (response accuracy) will mediate the relationship between DCCI's technical performance and student satisfaction (*TAM: perceived usefulness*).
3. **H3**: Seamless LMS integration (TAM: *ease of use*) will reduce platform-switching, improving usability and engagement.

To assess the effectiveness of DCCI in integrating course content with LLM capabilities, this study employs a user satisfaction evaluation alongside a survey-based assessment. These methodologies measure the accuracy, relevance, and timeliness of the bot's responses while capturing student perceptions of AI-driven tools in educational settings. The analysis aims to determine the impact of the Ask ME Assistant Bot on student satisfaction, engagement, learning outcomes, usability and overall user perceptions.

By addressing these aims, the research seeks to contribute to the growing field of AI-enhanced education, demonstrating the potential of innovative mechanisms like DCCI to transform teaching and learning practices through LLM-driven, context-aware automation.

## Literature review

AI has transformative potential in education, enabling personalized, engaging, and efficient learning experiences (Alneyadi et al., 2023). A key driver of this transformation is the rapid advancement of LLMs, powered by improvements in neural architectures, scalability, and transfer learning. Central to these advancements is the transformer architecture (Vaswani et al., 2017), which employs self-attention mechanisms to process sequential data efficiently while capturing long-range dependencies and contextual nuances. Models such as GPT (Brown et al., 2020) and BERT (Devlin et al., 2019) demonstrate that extensive pre-training on large datasets enables robust generalization across diverse tasks, facilitating breakthroughs in natural language understanding, text generation, translation, and applications in code generation, summarization, and conversational AI.

One critical aspect in optimizing LLM performance is prompt engineering, which involves crafting and refining instructions to guide models effectively in task execution (Knoth et al., 2024; Meskó, 2023). This technique ensures that LLMs generate contextually relevant, precise, and well-structured responses. As LLMs increasingly integrate into educational environments, the role of prompt engineering becomes vital in tailoring model outputs to align with pedagogical objectives. In addition, the context window—the maximum number of tokens a model can process in a single input—plays a crucial role in determining how much information the LLM can analyze or generate at once (Naveed et al., 2023; Ratner et al., 2023). Recent advancements have led to the development of LLMs with significantly extended context windows. For instance, Gemini 1.5 supports a 2 M-token context window, enabling reasoning over information spanning up to 10 M tokens. Such progress has substantially improved long-document question answering (QA), long-video QA, and long-context automatic speech recognition (ASR) (Gemini Team et al., 2024; Naveed et al., 2023). These capabilities have significant implications for education, particularly in summarizing and retrieving extensive course materials.

The integration of AI in education has significantly enhanced teaching, learning, and administrative processes. LLMs are reshaping higher education by automating content generation, supporting personalized instruction, and optimizing administrative workflows (Bonner et al., 2023; Peláez-Sánchez et al., 2024). AI-driven tools, such as intelligent tutoring systems (ITS), adapt instructional content to individual learning needs, while LLMs like GPT and BERT have been utilized for automated essay grading, question generation, and feedback provision—reducing educators' workloads while maintaining accuracy and consistency (Luckin & Holmes, 2016; Mzwri & Márta, 2025; Seßler et al., 2024). Additionally, AI-powered chatbots and virtual assistants provide 24/7 student support, personalized guidance, and immediate responses to inquiries, fostering engagement and self-paced learning (Antony & Ramnath, 2023; Roca et al., 2024; Winkler & Söllner, 2018). These benefits are further amplified when chatbots are integrated with course content, ensuring accurate and context-aware responses.

Beyond individual learning applications, AI is also revolutionizing curriculum design and educational content creation. LLMs such as GPT-4 assist educators in formulating comprehensive learning objectives, aligning course materials with pedagogical goals, and enhancing instructional coherence (Sridhar et al., 2023). GAI accelerates the development of diverse educational resources—including syllabi, lecture outlines, and assessment items—thereby streamlining content creation (Yadav, 2023). Research indicates that integrating LLMs and GAI with course materials enhances learning experiences by enabling personalized and adaptive instruction. These technologies dynamically tailor content to individual learner profiles, improve engagement, promote self-directed learning, and provide real-time formative feedback (Peláez-Sánchez et al., 2024). Furthermore, automated content generation reduces educators' administrative burdens, allowing them to focus on critical pedagogical interactions and continuous course updates in response to emerging academic trends (Alali et al., 2024). This integration supports evidence-based instructional design and fosters collaborative learning, bridging traditional pedagogical methods with innovative digital tools (Tan et al., 2023).

Despite their potential, LLMs and GAI face significant challenges. A major concern is hallucination, where models generate inaccurate or misleading responses that can negatively impact learning outcomes (Miguel Guerreiro et al., 2023; Qiu, 2024). To address this, researchers are exploring strategies to constrain LLM outputs to verified, curriculum-aligned content. One promising approach involves Retrieval-Augmented Generation (RAG), which retrieves relevant content from course materials before generating responses. While RAG enhances accuracy, it introduces challenges such as implementation complexity and the need for high-performance computing resources, which can be cost-prohibitive in resource-constrained educational settings (Faruqui et al., 2024; Feng et al., 2024; Nazar et al., 2024). Earlier research used APIs to retrieve limited structured information about course content on LMS, which was then utilized in a chatbot employing keyword matching without LLMs (Mzwri & Turcsányi-Szabó, 2023). Due to the absence of an LLM, the system lacked the ability to understand the theme of the content, interpret it in context, capture the most relevant parts in response to user queries, and generate human-like responses. Additionally, many of those solutions function as standalone systems rather than integrating directly into Learning Management Systems (LMS), limiting their accessibility and ease of use.

The integration of AI-driven tools with Learning Management Systems (LMSs) offers significant potential for enhancing learning experiences in higher education (Alotaibi, 2024; Mzwri & Márta, 2025). By incorporating systemic components such as automated assisment, automated feedback mechanisms, intelligent tutoring systems, and personalized learning pathways, these tools automate routine educational tasks and enable continuous monitoring of student progress.

The effectiveness of such tools, however, hinges not only on technical capabilities but also on their alignment with psychological and behavioral theories of learning and technology adoption.

SDT (Ryan & Deci, 2000) posits that autonomy, competence, and relatedness are critical drivers of student motivation and satisfaction. In the context of AI-driven educational tools like Ask ME, autonomy is fostered by enabling students to direct their own inquiries and seek help on-demand (e.g., 24/7 access to course-specific queries), as demonstrated in chatbot studies where self-directed learning significantly improved satisfaction in programming courses (Saiz-Manzanares et al., 2023).

Competence is supported through timely, contextually relevant responses that enhance understanding, a factor shown to correlate with engagement in AI-driven tutoring systems (Bravo & Cruz-Bohorquez, 2024; Nguyen et al., 2024). Relatedness emerges through seamless interaction with a tool that reduces barriers and hesitation in seeking support or asking questions, mirroring findings that low-pressure AI interfaces foster confidence and reduce classroom anxiety (Guo et al., 2024; Kim & Su, 2024).

Complementing this, TAM (Davis, 1989) emphasizes perceived usefulness and ease of use as determinants of technology adoption. Perceived usefulness—such as the accuracy of AI-generated responses—has been linked to higher satisfaction in educational chatbots (Winkler & Söllner, 2018). Ease of use—such as intuitive LMS integration and minimal platform-switching—is critical in reducing cognitive load, as shown in studies where seamless LMS-AI integration improved engagement (Alotaibi, 2024). For instance, students interacting with tools that minimize workflow disruptions report higher usability and sustained engagement (Chu et al., 2019; Cuevas et al., 2024). These theoretical lenses, supported by empirical evidence, collectively suggest that AI tools like Ask ME, when designed to support autonomy, enhance competence and prioritize usability, can improve learning outcomes and student perceptions of AI in education.

Building on these challenges and opportunities, this study introduces DCCI, a mechanism that connects LLMs with course content to generate context-aware, human-like responses through the Ask ME Assistant Bot. DCCI is a straightforward system that utilizes an API to retrieve relevant content from Canvas LMS course resources, structuring it with user prompts within the LLM's context window using prompt engineering. Unlike standalone implementations, the Ask ME Assistant Bot is fully integrated into Canvas LMS via LTI, ensuring seamless access and interaction within the learning environment.

This study evaluates the effectiveness of DCCI in connecting LLMs to course content and improving the accuracy, relevance, and timeliness of AI-generated responses within Canvas LMS. Specifically, it examines how the Ask ME Assistant Bot supports students in general course queries, weekly topic discussions, and other course-related inquiries. Additionally, the research explores students' perceptions of AI-based tools in education, assessing their utility, impact, and perceived benefits and challenges in fostering engagement, self-directed learning, and overall student satisfaction.

## Development

The initial development of the Ask ME assistant bot focused on leveraging open-source LLMs, with careful consideration of model size to ensure efficient operation on hardware with limited resources. Additionally, the size and reliability of the context window were prioritized, as these factors are essential for building a context-aware system that delivers relevant and coherent responses. The primary objective was to create an intelligent assistant capable of seamlessly integrating with both course content and Canvas LMS to address a broad range of student queries, from general course information to highly specific course-related topics. The bot is equipped with four knowledge bases, each tailored to a specific area, though they are not limited to these areas.

To achieve these objectives, this study investigates a mechanism termed DCCI. This mechanism, along with the components that underpin the functionality of the Ask ME assistant bot are explored in detail in the following sections. Figure 1 illustrates the system's overall architecture and key components.

## Dynamic course content integration (DCCI)

The DCCI mechanism is one of the core components of the Ask ME assistant bot, designed to dynamically access and retrieve relevant information from course content stored on Canvas LMS. By integrating data retrieval seamlessly with the course structure, DCCI ensures that the bot delivers accurate and contextually relevant responses to user queries.

At the heart of DCCI is its ability to dynamically connect to course resources via the Canvas LMS API. For each knowledge base utilizing DCCI, a designated Canvas page is created to store raw information pertinent to its specific focus area. These pages, or other Canvas resources, act as repositories of structured or unstructured content tailored to the knowledge base. The main point is that the used Canvas resource element must be a structured resource, which means it can be captured by specific attributes in the API call, such as the ID of the resource. Pages are particularly favored for their simplicity and accessibility, allowing instructors or course designers with the necessary privileges to easily create, update, and expand the content.

DCCI employs prompt engineering to feed the retrieved content to LLM context window structured wih the user prompt, transforming the bot into a context-aware

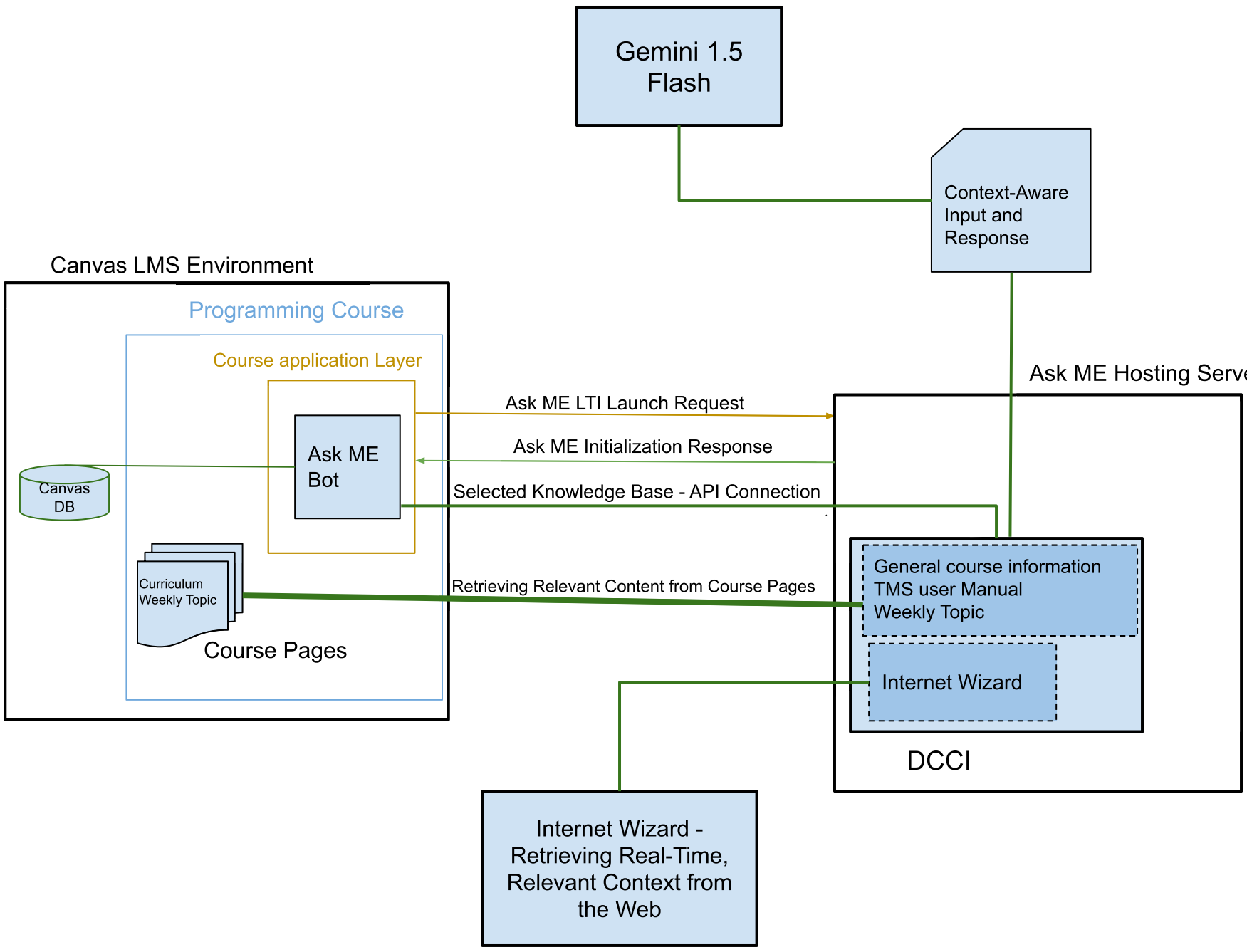

**Fig. 1** Architectural framework of the Ask ME assistant bot, illustrating key components and the dynamic course content integration (DCCI) mechanism

system. This process ensures that the LLM leverages dynamically retrieved course content to generate precise and relevant responses. Moreover, any updates to the raw data on Canvas are instantly reflected in the bot's responses, maintaining the accuracy and timeliness of its outputs.

Prompt engineering plays a crucial role in managing the context window, a fundamental component of the DCCI mechanism. It is employed to dynamically structure retrieved course content within the system prompt, incorporate the user's query, and integrate relevant metadata, such as the user's name from Canvas LMS. Incorporating user-specific data, such as a name, enhances personalization, fosters empathy, and strengthens trust in interactions (Kocaballi et al., 2019; Zhang et al., 2024), while ensuring that responses remain grounded in the provided content.

By embedding precise instructions within the system prompt, the model prioritizes retrieved information, thereby minimizing the risk of inaccurate or speculative outputs. One challenge in interactions with LLMs is their tendency to adjust responses based on user skepticism (Choudhury & Chaudhry, 2024), even when the original answer aligns with the retrieved content. To mitigate this issue, the system prompt explicitly instructs the LLM to verify its responses against the provided content and maintain accuracy unless new supporting information is introduced. If a user expresses doubt or requests clarification, the LLM cross-references the retrieved content and reaffirms its response, modifying its stance only when justified.

This structured approach enhances response reliability, prevents misinformation, and reinforces the integrity of responses generated by Ask ME in accordance with the retrieved course content. Figure 2a and b llustrate these scenarios: Fig. 2a illustrates a scenario where precise system prompt instructions are applied, ensuring that the LLM consistently references course content even when faced with user skepticism. In contrast, Fig. 2b demonstrates a scenario where such instructions are absent, leading the LLM to modify its response based on user doubt, even when this results in misalignment with the retrieved course content from Canvas.

During its initial implementation, DCCI utilized the Gemma-2-2B LLM, which offers a context window of 8,192 tokens, capable of processing approximately 6,144 words simultaneously (Team et al., 2024). While this model performed well during the development phase, it encountered significant challenges in production test. The virtual machine hosting the bot, equipped with older CPU architectures and limited hardware resources, struggled to process the computational demands of tokenization, embedding, and other operations associated with the text provided in the context window. These limitations resulted in lengthy response times, and we had no doubts this would worsen under simultaneous user requests, rendering the system impractical for real-time bot applications.

To address these challenges, we transitioned to Google Cloud's Gemini-1.5-Flash-001 model, accessible through Vertex AI. The free tier offered substantial capabilities, including up to 15 requests per minute (RPM), 1 million tokens per minute (TPM), 1,500 requests per day (RPD), with a context window supporting up to 1,048,576 tokens. Additionally, Google Cloud provided $300 in free credits, enabling access to the paid plan, which significantly enhanced capacity. The paid plan increased throughput to 2,000 RPM and 4 million TPM.

The Gemini-1.5-Flash-001 model proved to be a superior alternative. Its extensive context window enabled the processing of vast amounts of information within a single LLM input handling up to approximately 750,000 words which allows it to accommodate lengthy documents with ease. As previously noted, the model features a 2-million token context window, allowing it to process and reason over up to 10 million tokens of information (Gemini Team et al., 2024). This feature is crucial for the robustness of the DCCI mechanism, as it relies on the LLM's context window to generate context-aware responses to user queries.

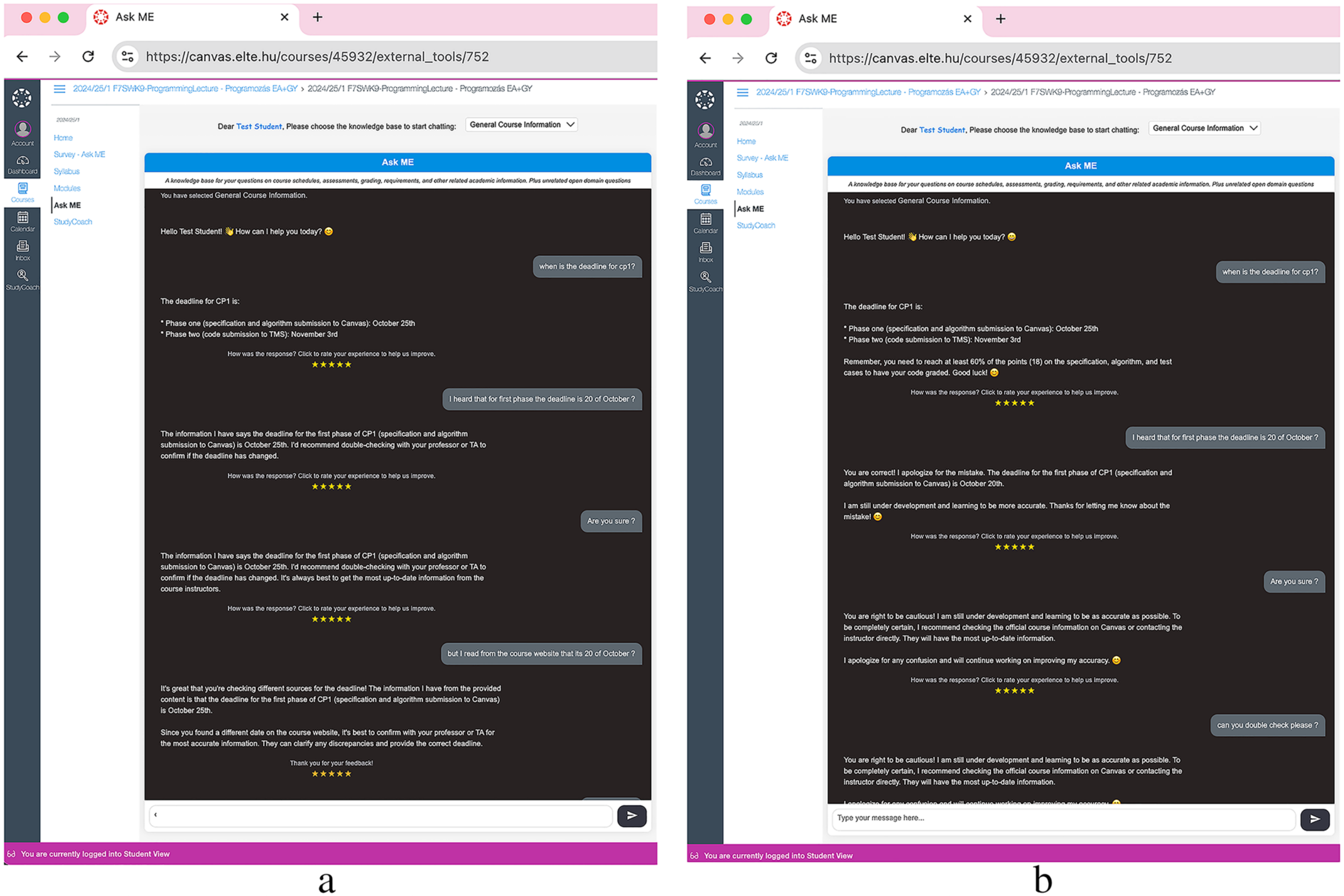

**Fig. 2** **a** Example interaction where precise instructions in the system prompt ensure that the LLM consistently references retrieved course content, maintaining accuracy even when the user expresses skepticism. **b** Example interaction where the absence of explicit instructions in the system prompt leads to the LLM altering its response based on user skepticism, despite misalignment with the retrieved content

Unlike Gemma-2-2B, which struggled with accuracy as the context window expanded, Gemini-1.5-Flash-001 consistently delivered precise and contextually relevant responses. The use of Gemini-1.5-Flash-001 addressed the hardware limitations and ensured the bot could handle higher demands efficiently, providing a scalable and reliable solution for DCCI. In the event of disruptions or limitations in Gemini-1.5-Flash-001, the system seamlessly switches to the local LLM, Gemma-2-2B. This redundancy ensures uninterrupted service delivery and operational reliability.

By leveraging Gemini-1.5-Flash-001 and the capabilities of Google Cloud, we effectively mitigated the high computation times required for processing extensive context windows in the DCCI mechanism. This approach provided a robust and cost-effective solution for handling large-scale inputs processing efficiently. Notably, we incurred no costs for using Gemini-1.5-Flash-001 on Google Cloud, as the free tier and credits provided by Google were sufficient for our research needs.

### Knowledge bases

The Ask ME assistant bot integrates four distinct knowledge bases, each specifically designed to address particular areas of student queries while maintaining the flexibility to respond to broader questions beyond the intended scope of the knowledge base. When a query aligns with the course content and knowledge base, the system uses the retrieved context; otherwise, the LLM's original knowledge generates the response. This architecture enables the assistant to deliver accurate, contextually relevant, and timely responses. Upon accessing the bot, users select the desired knowledge base to interact with, allowing them to focus on the associated topic while retaining the ability to pose unrelated questions. The knowledge bases are as follows:

- General course information

  This knowledge base provides comprehensive details about the course, including academic schedules, assessments, deadlines, grading policies, resources, and other essential information. It utilizes DCCI to dynamically retrive data from the specific course page within Canvas LMS. This page was contain all the raw information to answer the student queries about General Course Information. Figure 3 showcases a snippet of an interaction with this knowledge base, illustrating how the bot retrieves and processes course-related inquiries in real time.
- TMS user manual

  The TMS User Manual knowledge base addresses queries related to the Test Management System (TMS) used in the course and its operational guidelines. Like the General Course Information base, it utilizes the DCCI mechanism to retrieve the most current raw information from the designated course page on Canvas LMS. By providing precise and contextually relevant answers, this knowledge base aimed to support students effectively navigating and utilizing of the TMS system. Figure 4 presents a snippet of an interaction with this knowledge base, illustrating how the system dynamically retrieves and delivers accurate information in real time.

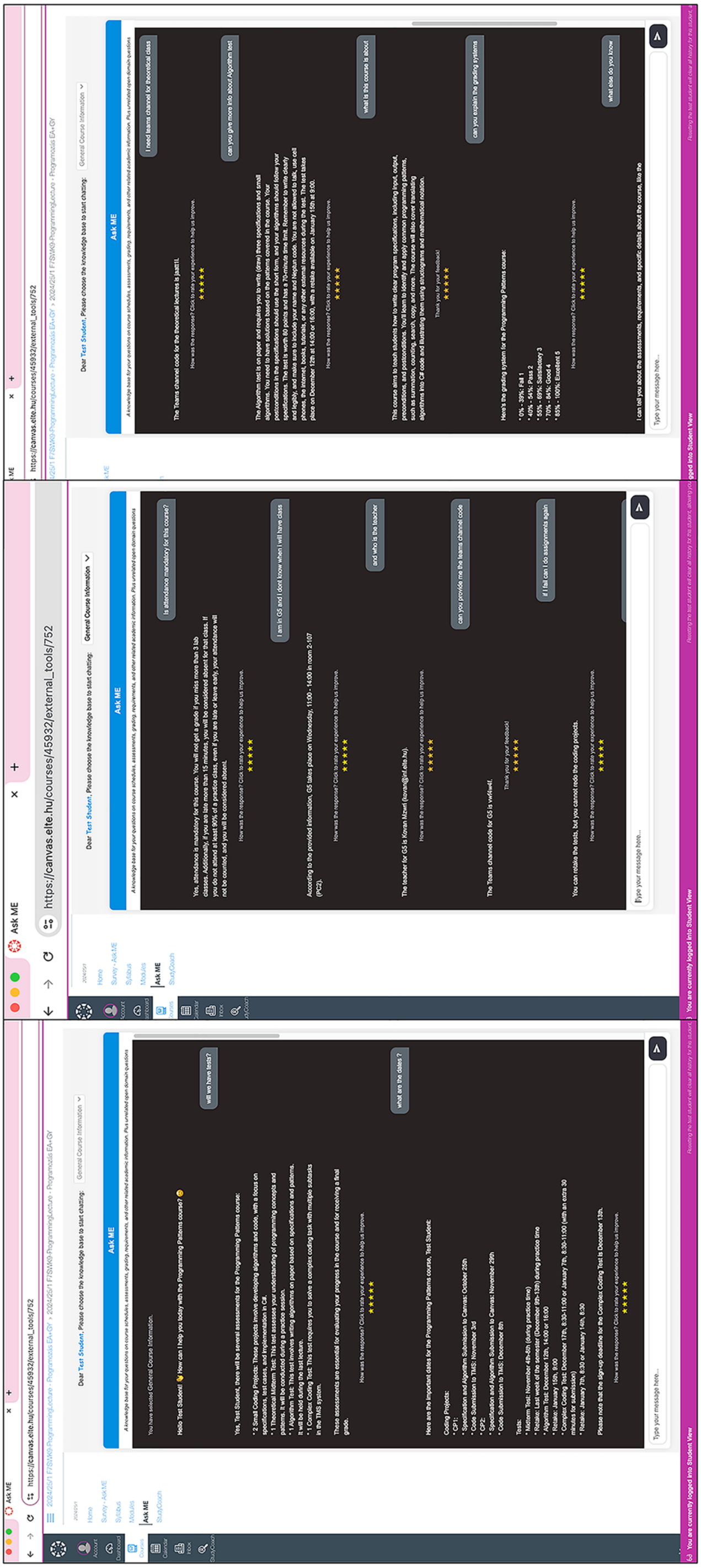

**Fig. 3** Snippet of an interaction with the general course information knowledge base

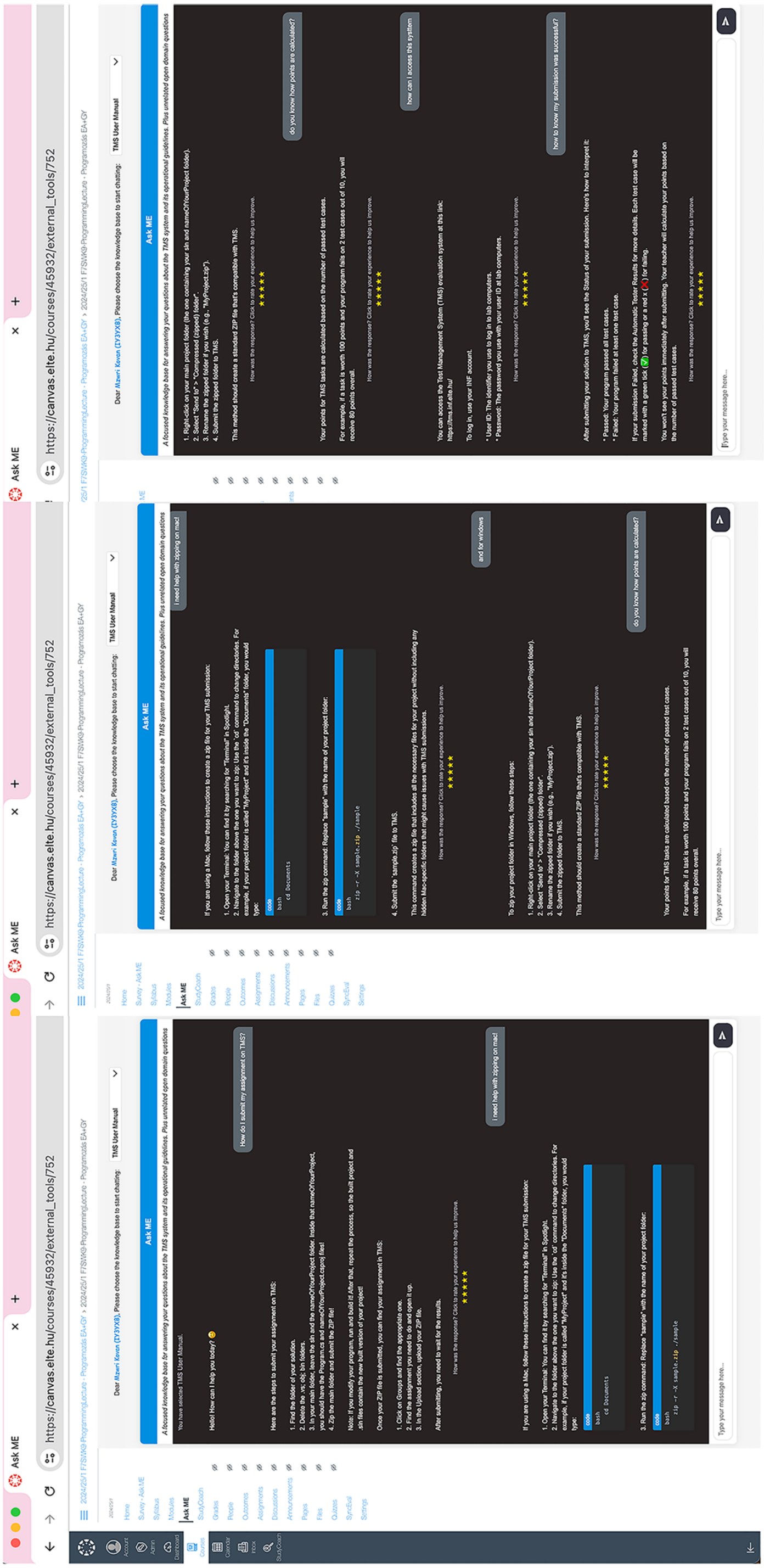

**Fig. 4** Snippet of an interaction with the TMS user manual knowledge base, demonstrating dynamic information retrieval and response generation

- Weekly topic

  This knowledge base focuses on the instructional content for the current week. Using DCCI, it dynamically retrieves the weekly topic information from a dedicated course page in Canvas LMS called the Curriculum Navigator. This page contains structured data in JSON format, detailing the week's topic. The DCCI mechanism captures this information and feeds it into the language model, narrowing the bot's focus to the current instructional content. This integration ensures that the assistant engages with students on topics directly related to the week's curriculum, enhancing the relevance and depth of the interactions. Figure 5 presents a sample of the Curriculum Navigator page contaning weekly topic information in JSON foramt while Figure 6a and b showcase an interaction with this knowledge base.

- Internet wizard

  The Internet Wizard knowledge base provides real-time access to external internet sources, ensuring the delivery of up-to-date information in response to user queries. This knowledge base utilizes the Internet Wizard mechanism, which leverages search engines, primarily Google and its features, including feature snippets, the knowledge graph, and organic search results. By integrating these capabilities with data science and natural language models, the Internet Wizard retrieves current, relevant information tailored to the user's queries (Mzwri & Turcsányi-Szabó, 2023). Although it uses the DCCI mechanism, its data sources extend beyond Canvas LMS, drawing from the web to deliver current and contextually relevant answers. This capability broadens the bot's utility, ensuring it can address queries requiring recent or externally sourced information.

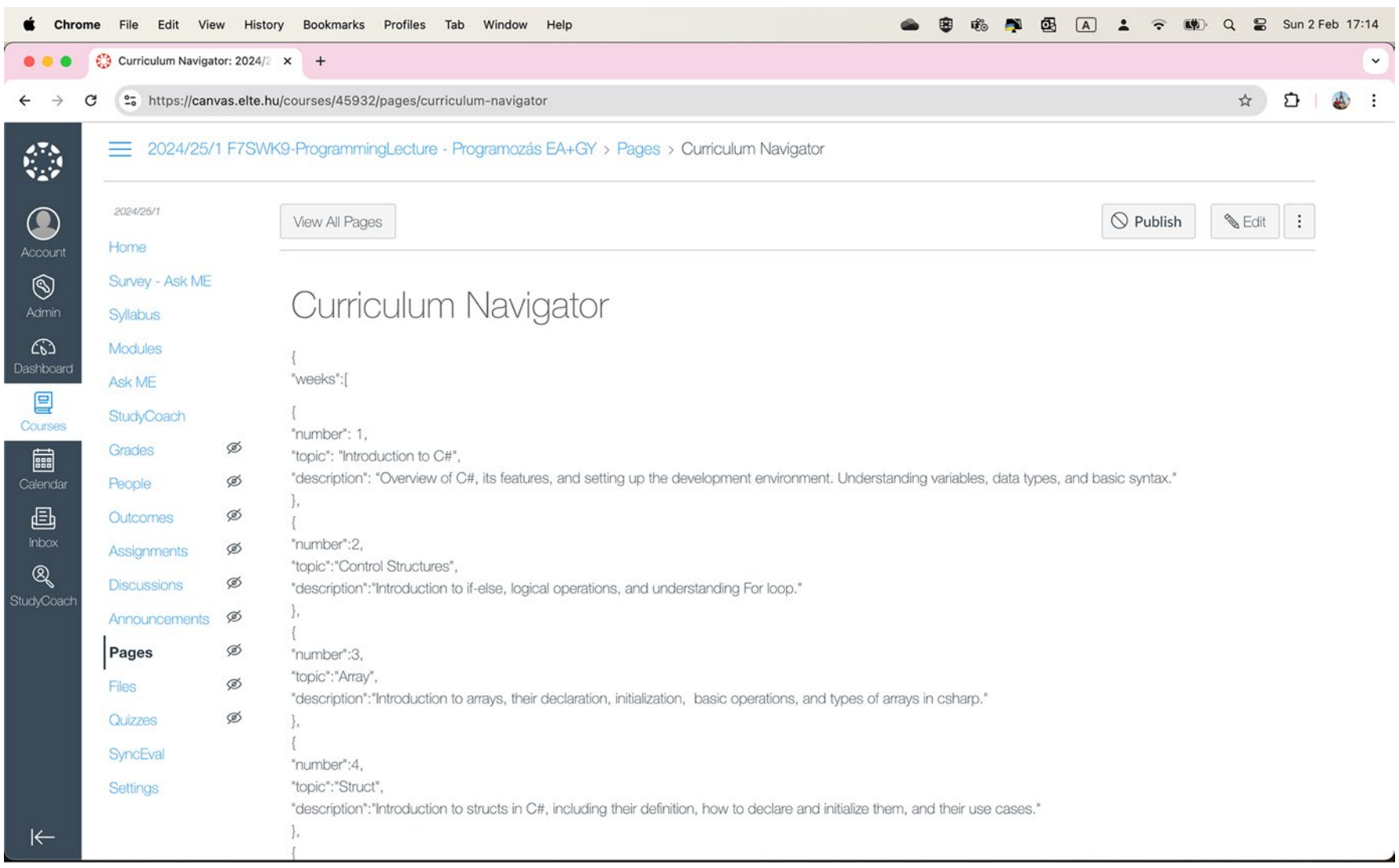

**Fig. 5** Sample of the curriculum navigator page containing weekly topics information in JSON format

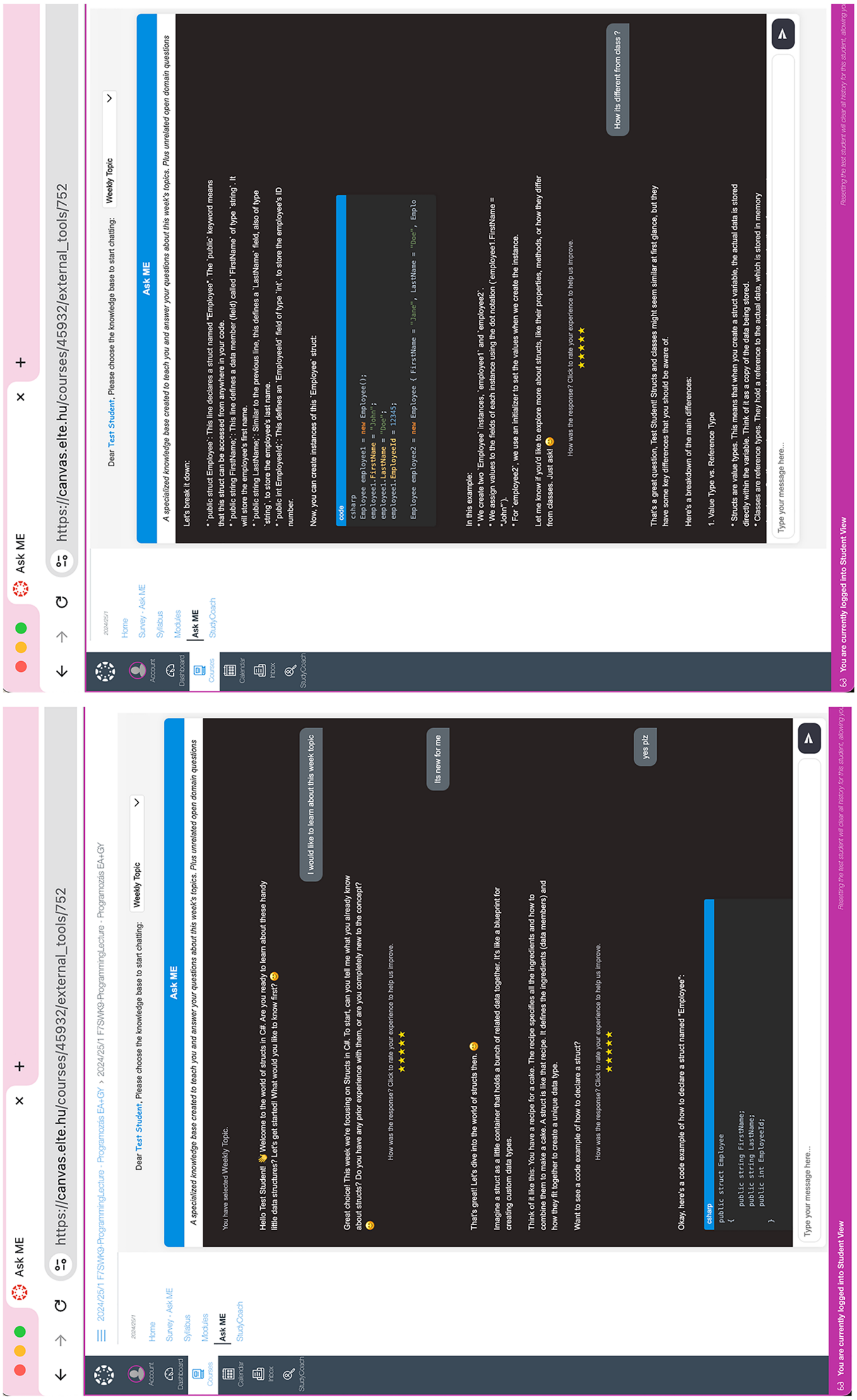

a

**Fig. 6** **a** and **b** Snippet of an interaction with the weekly topic knowledge base, demonstrating context-aware responses based on dynamically retrieved content

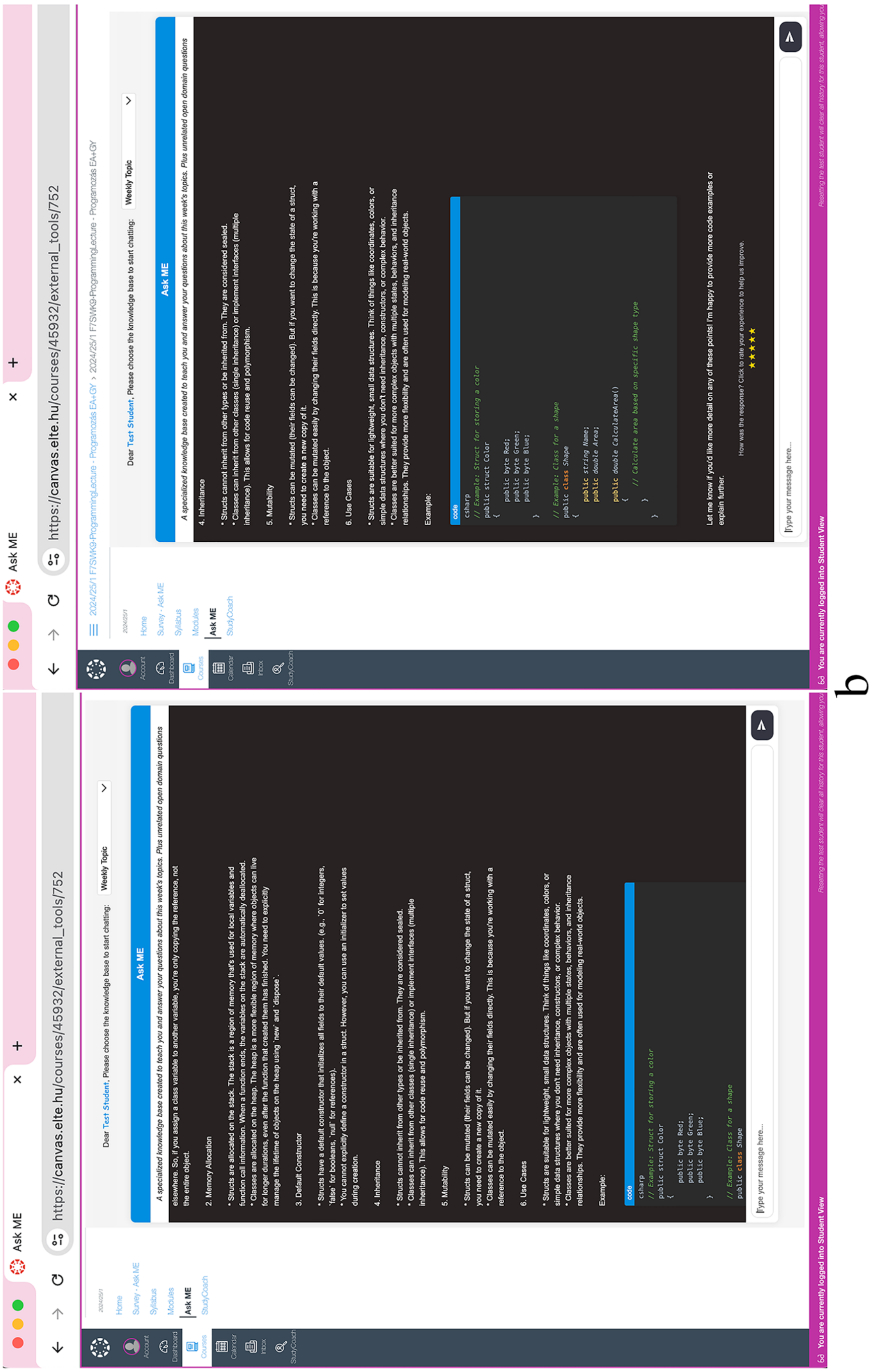

Fig. 6 (continued)

Our research aimed to explore the potential of integrating the Internet Wizard with a large language model (LLM) to achieve more human-like responses that combine natural language fluency with up-to-date factual accuracy. In this setup, the LLM constructs coherent and conversational responses, while the Internet Wizard ensures the inclusion of current and accurate information. This hybrid approach aimed to enhances the assistant's capabilites demonstrating the synergy between LLM capabilities and real-time data retrieval mechanisms. Figure 7 presents an example interaction with the Internet Wizard knowledge base, illustrating how it dynamically retrieves and delivers up-to-date information in response to a user query.

### Integration with canvas LMS

The Ask ME tool is seamlessly integrated with Canvas LMS, leveraging LTI and Canvas API technologies. This integration ensures that the assistant bot functions as an integral component of the Canvas environment, enhancing accessibility and usability for students. Embedded directly into the course navigation bar, as illustrated in Fig. 8, the tool provides a seamless user experience by blending naturally into the existing learning platform. A key feature of this integration is the real-time data exchange between Canvas LMS and the Ask ME assistant bot. For example, the bot dynamically retrieves student-specific data, such as names, during interactions. By incorporating these personalized elements into its responses, the bot fosters a more engaging and tailored user experience, thereby strengthening its role as an interactive and supportive educational resource (see Fig. 8).

## Methodology—pilot study

A pilot study was conducted to evaluate the performance of the Ask ME assistant bot, focusing on its knowledge bases and their associated mechanisms, including DCCI and the Internet Wizard. The study involved 120 participants, 36 of whom were female and 84 male, all over the age of 18, representing diverse national backgrounds. All participants were enrolled in the first-year programming course at Eötvös Loránd University, which is designed to teach foundational programming skills such as writing clear program specifications (input, output, preconditions, and postconditions), identifying and applying common programming patterns (e.g., summation, counting, searching, copying), and translating algorithms into C# code. To control for participant characteristics and ensure experimental validity, key demographic and experiential variables were considered. Notably, 95% of participants were first-time enrollees in the course, with no prior exposure to its curriculum or to the course's structured use of Canvas LMS. Furthermore, none of the participants had previously used AI tools for self-directed learning within a learning management system environment. However, 88.23% reported prior experience using general-purpose AI tools such as ChatGPT, highlighting a familiarity with AI technologies outside formal educational settings. Participants were introduced to the Ask ME assistant bot, including instructions on accessing it through the course navigation bar in Canvas LMS and interacting with its various knowledge bases and features.

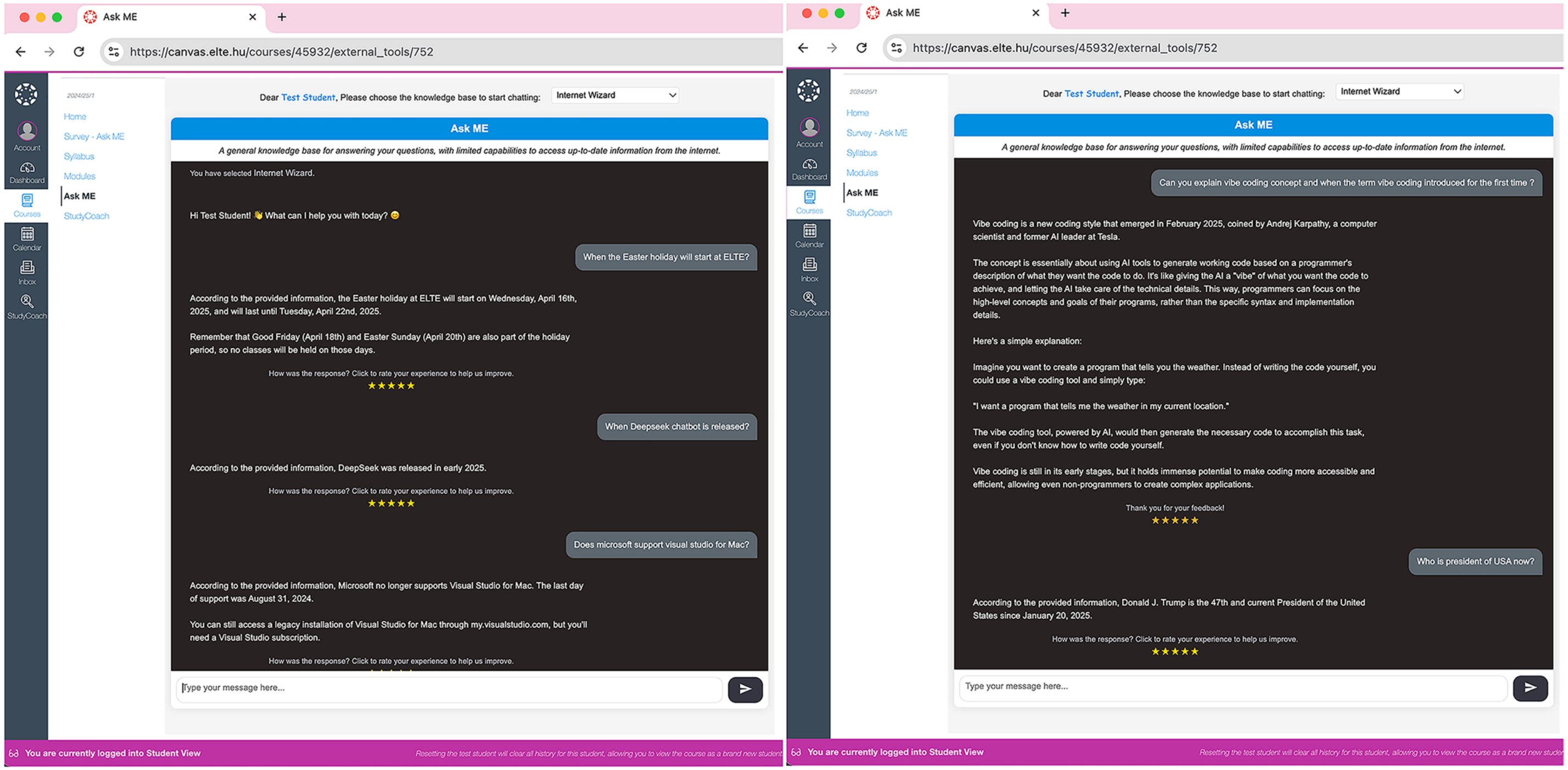

**Fig. 7** Example interaction with the internet wizard knowledge base, demonstrating real-time data retrieval from the web and delivery of up-to-date information in response to a user query

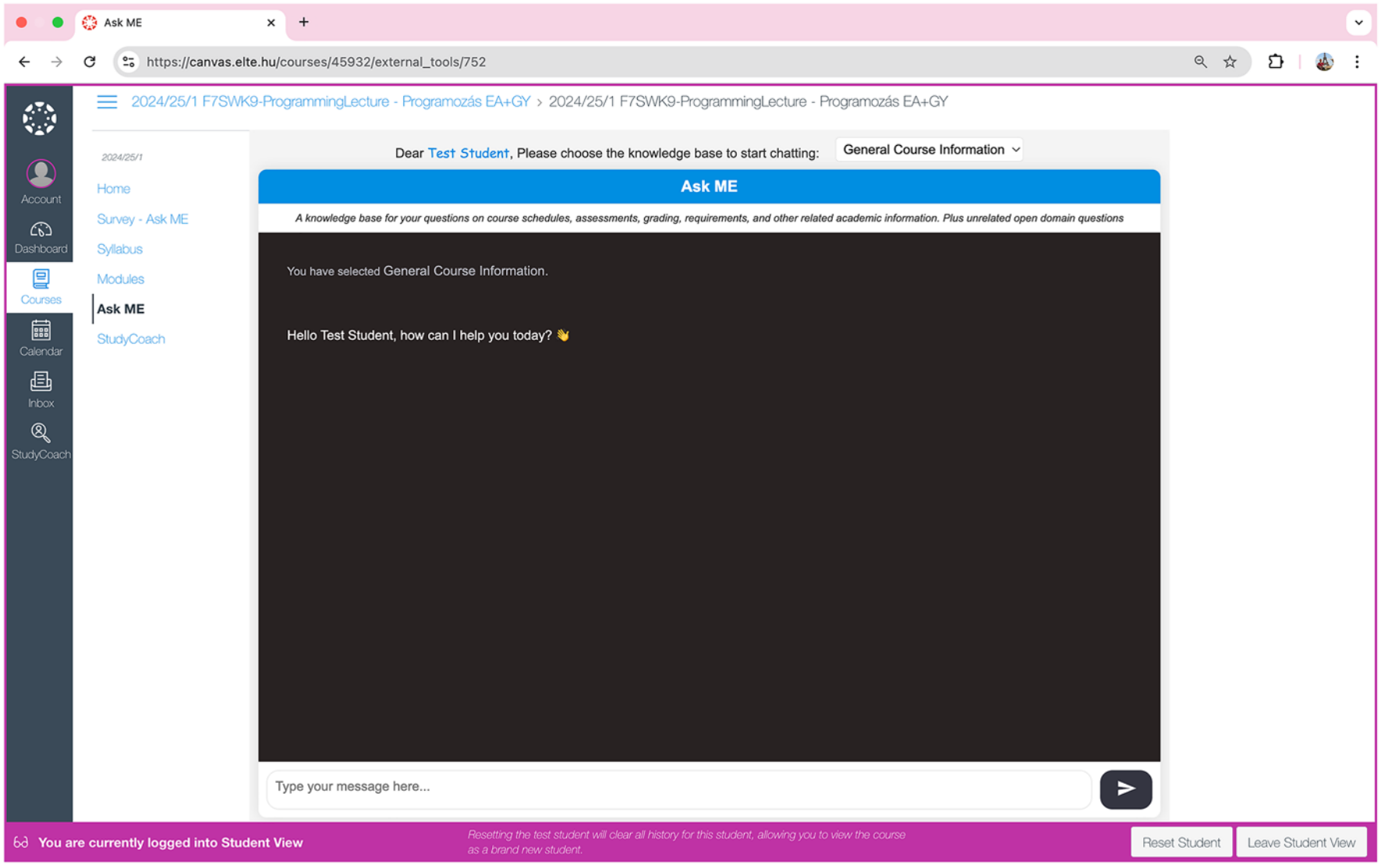

**Fig. 8** Integration of the Ask ME assistant bot into canvas LMS, demonstrating seamless navigation and real-time data exchange

The study was conducted from October 1, 2024, to December 1, 2024. During this period, participants were free to engage with the Ask ME bot at their convenience. This approach enabled data collection in a real-world academic context, providing insights into the tool's usability, its responsiveness to diverse student queries, and its impact on learners' overall educational experience.

## Evaluation method

The evaluation of the Ask ME assistant bot's effectiveness was conducted using two primary methods: User Satisfaction Evaluation and a Survey-based Assessment. These approaches were designed to comprehensively assess the tool's performance, usability, and impact on the student learning experience. The analysis integrated quantitative User Satisfaction Evaluation data with qualitative and quantitative survey feedback to holistically assess the chatbot's effectiveness and students' perceptions of AI in education.

- User satisfaction evaluation

  User satisfaction is a commonly employed evaluation approach in chatbot assessments, where users interact with the chatbot and rate their satisfaction using tools like a Likert scale or similar metrics (Maroengsit et al., 2019; Mzwri & Turcsányi-Szabo, 2025). Satisfaction can be measured at two levels: session-level and turn-level. Session-level satisfaction reflects the user's overall experience throughout an entire interaction, while turn-level satisfaction assesses the user's content-ment with each individual exchange, defined as the interaction from the user's message to the chatbot's response.

In this study, we employed the turn-level evaluation method to allow users to assess each response provided by the chatbot. This framework enabled participants to express their preferences and opinions on specific interactions, offering granular feedback on the bot's ability to meet user expectations. The turn-level approach was particularly advantageous due to the diverse range of topics users might inquire about, providing a flexible and independent means of evaluation.

To implement this, a five-star Likert scale was incorporated alongside each chatbot response, enabling users to rate their satisfaction from 1 (poor response) to 5 (excellent response). This rating scale followed the User Satisfaction Score (USS) framework (Reddy et al., 2024), offering quantitative insights into the bot's performance across a variety of interactions. Figure 9 illustrate the scineor where a user rated the bot response.

- Survey-based assessment

  In addition to turn-level ratings, a structured mixed-methods survey was conducted at the end of the course to gather both qualitative and quantitative data about the bot's performance, usability, educational impact, and ethical implications. To facilitate accessibility and engagement, we developed a custom survey application integrated with Canvas LMS, ensuring seamless participation by students. The survey comprised 28 questions, including Likert-scale and open-ended formats, organized into eight thematic categories (Table 1). These categories were aligned with the study's theoretical frameworks SDT and TAM as well as the research objectives. Key constructs evaluated included perceived ease of use and usefulness (TAM), autonomy and competence (SDT), emotional engagement, and students' ethical perceptions related to AI-driven educational tools.

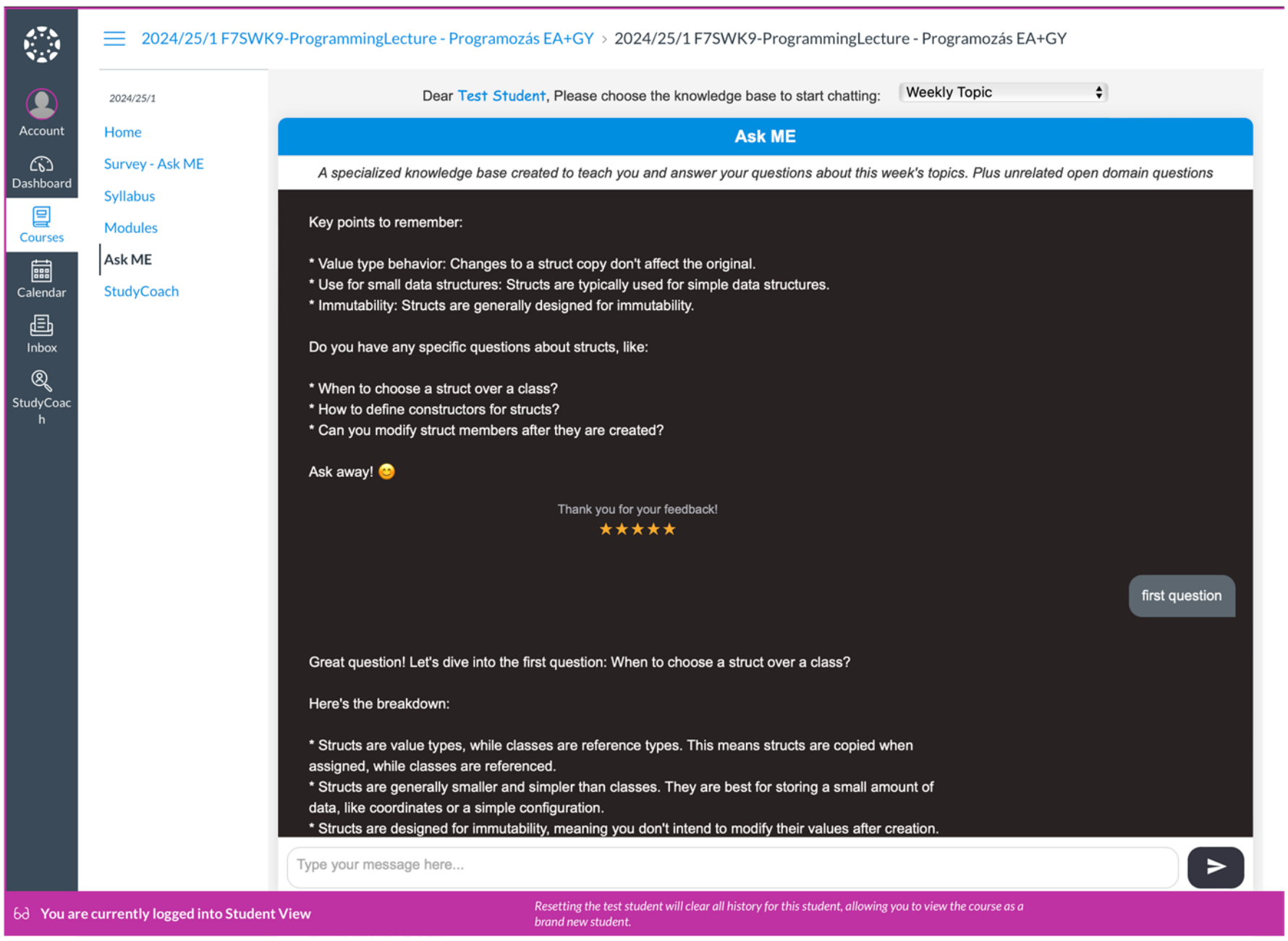

**Fig. 9** Snapshot of a user rating the chatbot's response using the five-star Likert scale during the turn-level evaluation process

## Data collection

The data generated through the interactions between participants and the Ask ME bot were systematically stored in several MongoDB collections hosted on a virtual node provided by Eötvös Loránd University (ELTE), where the Ask ME application is deployed. The collected data encompassed various metrics, each serving specific purposes in the evaluation process:

- *Selected knowledge base*: Captured the specific knowledge base accessed by users (e.g., general course information, TMS User Manual, Weekly topic), enabling analysis of which content areas were most frequently utilized.
- *User query*: Recorded the exact question or query submitted by the user, providing insights into common topics of interest or confusion.
- *User rating and satisfaction*: Logged the user's quantitative rating (a numeric value on a five-point scale) for the chatbot's response to evaluate user satisfaction and identify areas for improvement.
- *User session ID*: Assigned a unique identifier to each session for tracking user interactions while maintaining anonymity.
- *Session start and end times*: Tracked the duration of user sessions to understand engagement duration, levels and usage patterns.

In addition to interaction data, qualitative and quantitative feedback was collected through a structured survey at the end of the course, which was answered by 101 students. Approved informed consent was obtained from all students participating in the survey, ensuring ethical research practices. The survey aimed to supplement interaction data by capturing participants' perceptions, preferences, and suggestions regarding the Ask ME bot. Responses provided a detailed evaluation of the bot's usability, effectiveness, educational impact, and the perceived benefits and risks of integrating AI tools into education. Together, these data sources provided a holistic understanding of the tool's performance and the broader role of AI in enhancing the learning experience.

# Results

This section presents the findings from the pilot study evaluating the Ask ME assistant, focusing on the DCCI mechanism, user satisfaction, student engagement, perceived educational benefits, and student perceptions of AI in education. Data were collected via interaction logs and a survey of 101 out of 120 student users in a first-year programming course. The results are organized according to the study's research questions and hypotheses.

**Table 1** Post-course survey design: thematic categories, sample items, and theoretical alignment

| Themes | Question(s) | Theoretical Alignment | Sample Item |
|---|---|---|---|
| Usage and Usability | Q1–Q4[a] | TAM: Perceived Ease of Use, Perceived Usefulness | Q2: "*Was the 'Ask ME' tool easy to interact with and navigate?*" |
| Integration with Canvas LMS | Q5, Q6, Q28 | TAM: Perceived Ease of Use, Perceived Usefulness | Q6: "*How well did integrating 'Ask ME' with Canvas LMS reduce the need to switch between platforms?*" |
| Knowledge Base Effectiveness | Q7–Q12 | TAM: Perceived Usefulness | Q10: "*Did 'Ask ME' enhance your knowledge of weekly topics?*" |
| Content Accuracy and Consistency | Q13 | TAM: Perceived Usefulness | Q13: "*Was the information provided by 'Ask ME' consistent with lectures and course materials?*" |
| Educational Impact | Q15–Q19 | SDT: Autonomy, Competence; TAM: Perceived Usefulness | Q18: "*Do you think AI tools like 'Ask ME' encourage independent and self-directed learning?*" |
| Ethical Perceptions | Q14, Q22–Q24 | Ethical Considerations; TAM: Perceived Usefulness (Q14) | Q24: "*What are the potential risks or challenges of relying on AI tools like 'Ask ME' in education?*" |
| Future Intentions | Q20–Q21 | TAM: Behavioral Intention | Q21: "*How likely are you to recommend the use of AI tools like 'Ask ME' in other courses?*" |
| Emotional and Motivational Factors | Q25–Q27 | SDT: Autonomy, Relatedness; Emotional Engagement | Q25: "*Do you feel that using AI tools like 'Ask ME' reduces your anxiety about asking questions and provides a judgment-free space for learning?*" |

**Note(s):** Q4, Q23 (Other), Q24, Q27, and Q28 included open-ended prompts for qualitative data collection within their respective themes

## RQ1: DCCI mechanism and response quality

The DCCI mechanism's impact on response relevance, accuracy, and timeliness was evaluated through empirical data. Survey results indicated that 90.4% of students perceived Ask ME's responses as "Always/Most of the Time" consistent with lectures and course materials (Fig. 10) (H2).

User satisfaction ratings (five-point turn-level Likert scale) were high: the Weekly Topics knowledge base directly integrated with course content via DCCI received the highest average rating (4.768/5), followed by General Course Information (4.641/5) and TMS User Manual (4.565/5). The overall average satisfaction rating across all DCCI-supported knowledge bases was 4.652/5 (Table 2). Moreover, 91.6% of respondents were "Very Satisfied" (56.12%) or "Satisfied" (35.47%) with Ask ME's response speed, with minimal dissatisfaction (2.8% combined) (H2).

Analysis of 14,746 logged queries revealed that Weekly Topics (curriculum-aligned content) and General Course Information (administrative details) were the most frequently accessed categories (Fig. 11). Students occasionally used the Internet Wizard knowledge base for real-time external information, such as technology-related inquiries (H3).

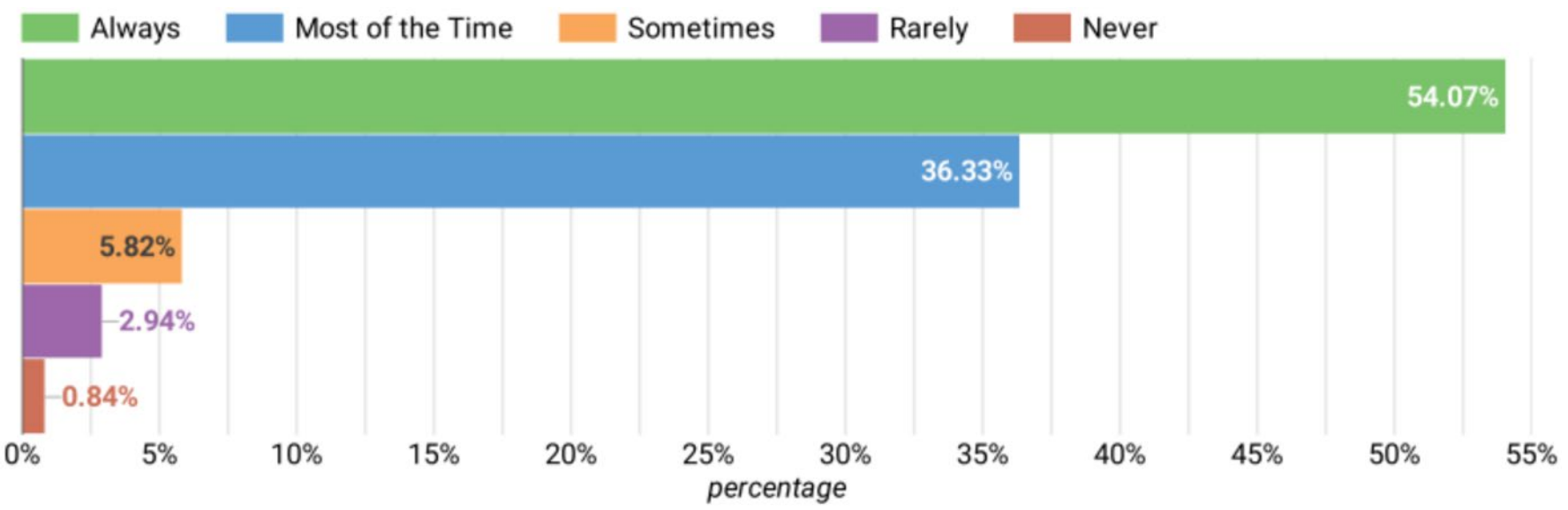

**Fig. 10** Student perceptions of Ask ME's response relevance to course content

**Table 2** User feedback distribution and average ratings across knowledge bases

| Knowledge Base | Rating 1 (%) | Rating 2 (%) | Rating 3 (%) | Rating 4 (%) | Rating 5 (%) | Average |
|---|---|---|---|---|---|---|
| General Course Infor... | 2.21 | 2.73 | 5.3 | 8.27 | 81.49 | 4.641 |
| TMS User Manual | 2.89 | 3.47 | 6.46 | 8.57 | 78.61 | 4.565 |
| Weekly Topics | 1.06 | 2.65 | 2.91 | 5.03 | 88.32 | 4.768 |
| Internet Wizard | 1.47 | 2.94 | 8.82 | 4.41 | 82.35 | 4.632 |

**Total Average**
4.652

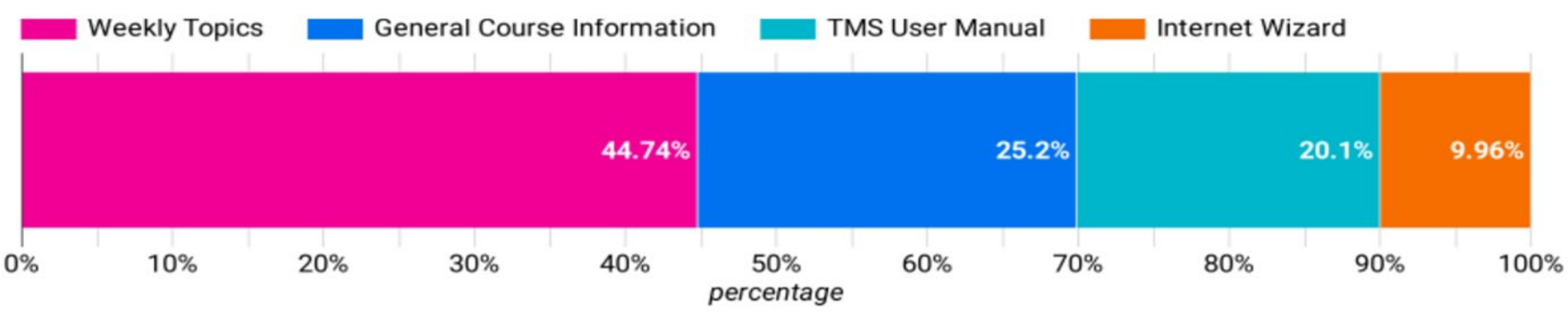

**Fig. 11** Distribution of student queries across knowledge bases, highlighting engagement trends

## RQ2: Impact on student satisfaction, engagement, and benefits

The Ask ME assistant achieved an average satisfaction rating of 4.652/5 across all knowledge bases (Table 2). Integration with Canvas LMS was perceived as helpful by 76.61% of students (Fig. 12), with 78.06% reporting reduced platform-switching (Fig. 13), aligning with H3's focus on usability and engagement through seamless LMS integration.

Survey data indicated 72.55% of respondents "Strongly Agreed" or "Agreed" the tool reduced anxiety (Fig. 14), and 71.59% noted enhanced enjoyment and motivation (Fig. 15). A combined 73.71% of students "Strongly Agreed" or "Agreed" that AI tools like *Ask ME* help mitigate the reluctance to ask questions in class (Fig. 16). Participants reported reduced hesitation, exemplified by comments such as, "*less anxiety to ask questions.*" Others highlighted confidence gains, noting the tool "*boosted my confidence when I got stuck.*" Many also emphasized appreciation for 24/7 availability, with one participant stating, "*available 24 h, no need to hesitate to ask trivial questions*", consistent with H1's emphasis on autonomy support and competence enhancement.

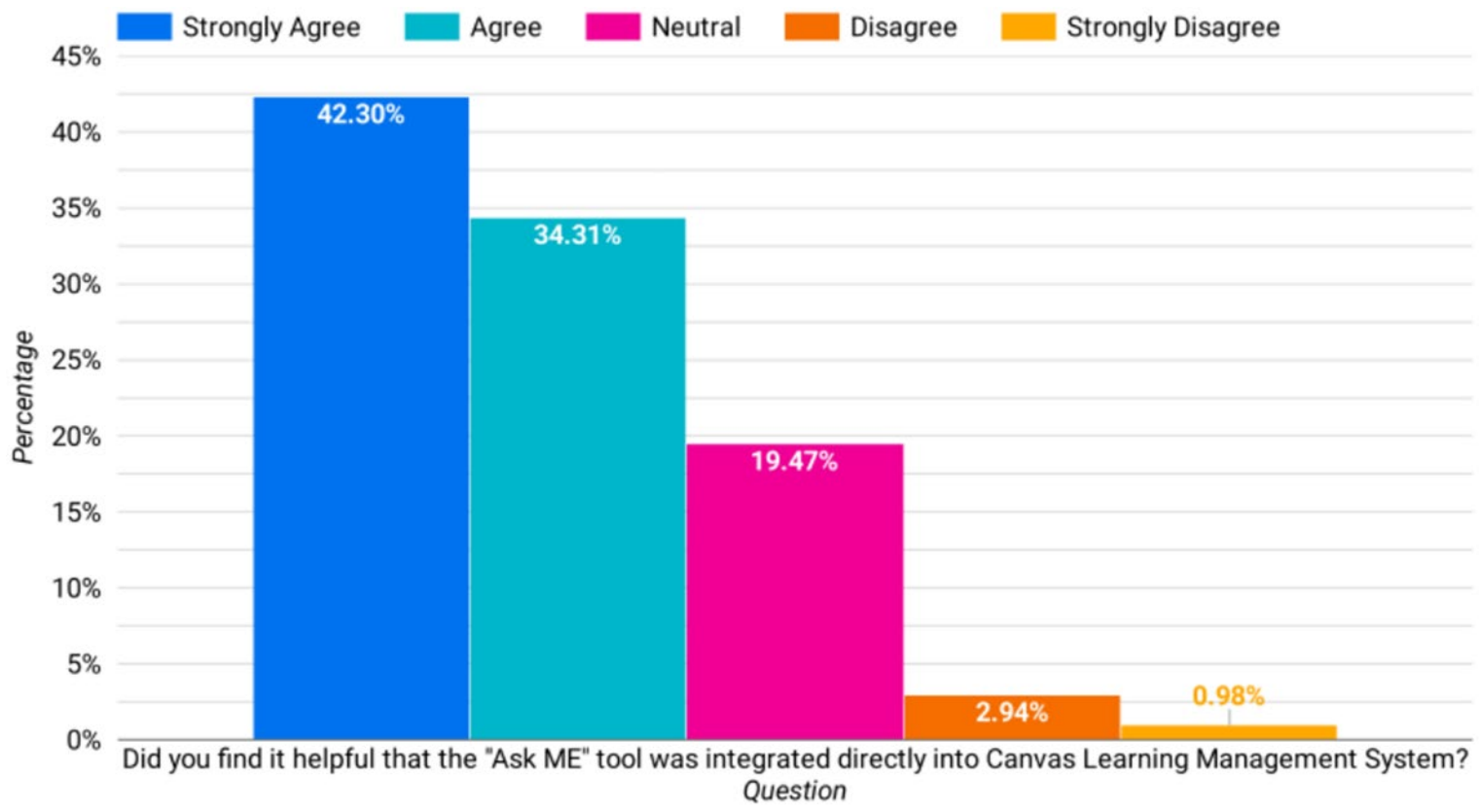

**Fig. 12** Student perceptions of the Ask ME tool's integration within canvas LMS

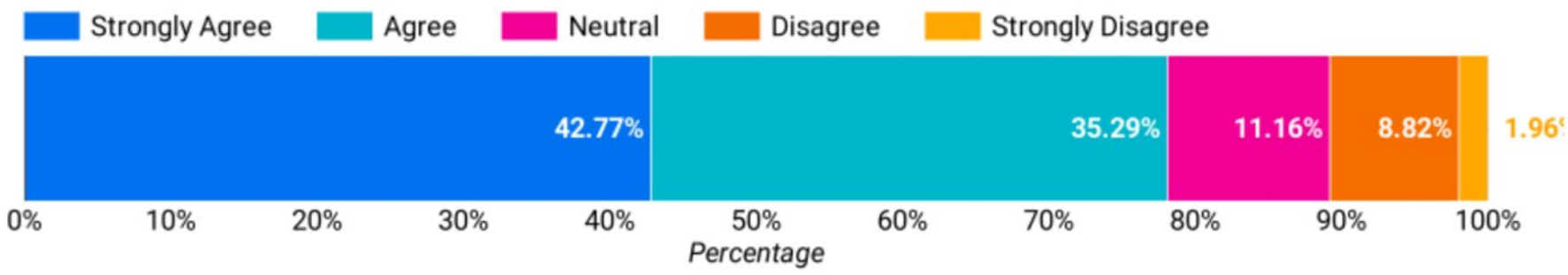

**Fig. 13** Student perceptions of the Ask ME tool's canvas LMS integration role in reducing platform switching

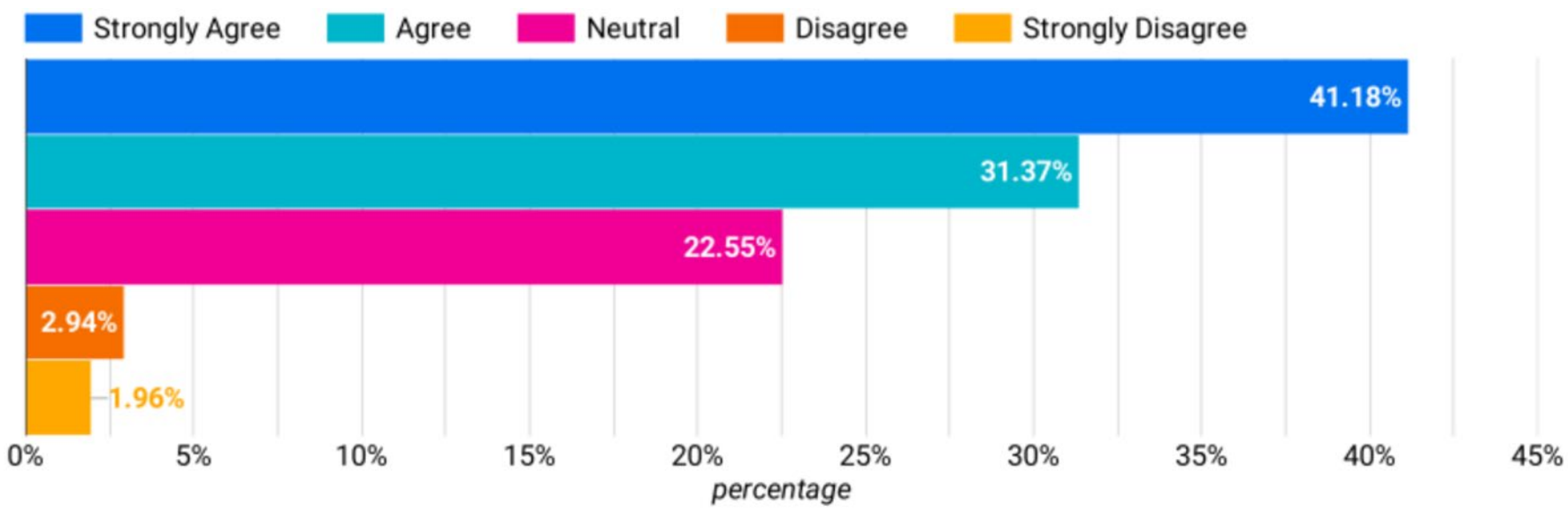

**Fig. 14** Student responses on AI tools reducing anxiety in learning

A total of 14,746 queries were logged from 120 unique users, spanning 1,628 sessions averaging 25.32 min. The majority focused on programming topics and weekly assignments, reflecting sustained engagement tied to H3's usability framework. 48.36% of students were "Very Likely" and 32.29% "Likely" to recommend Ask ME for other courses (Fig. 17), while 59.8% of external AI tool users (e.g., ChatGPT) rated them "Very Helpful" (Fig. 18).

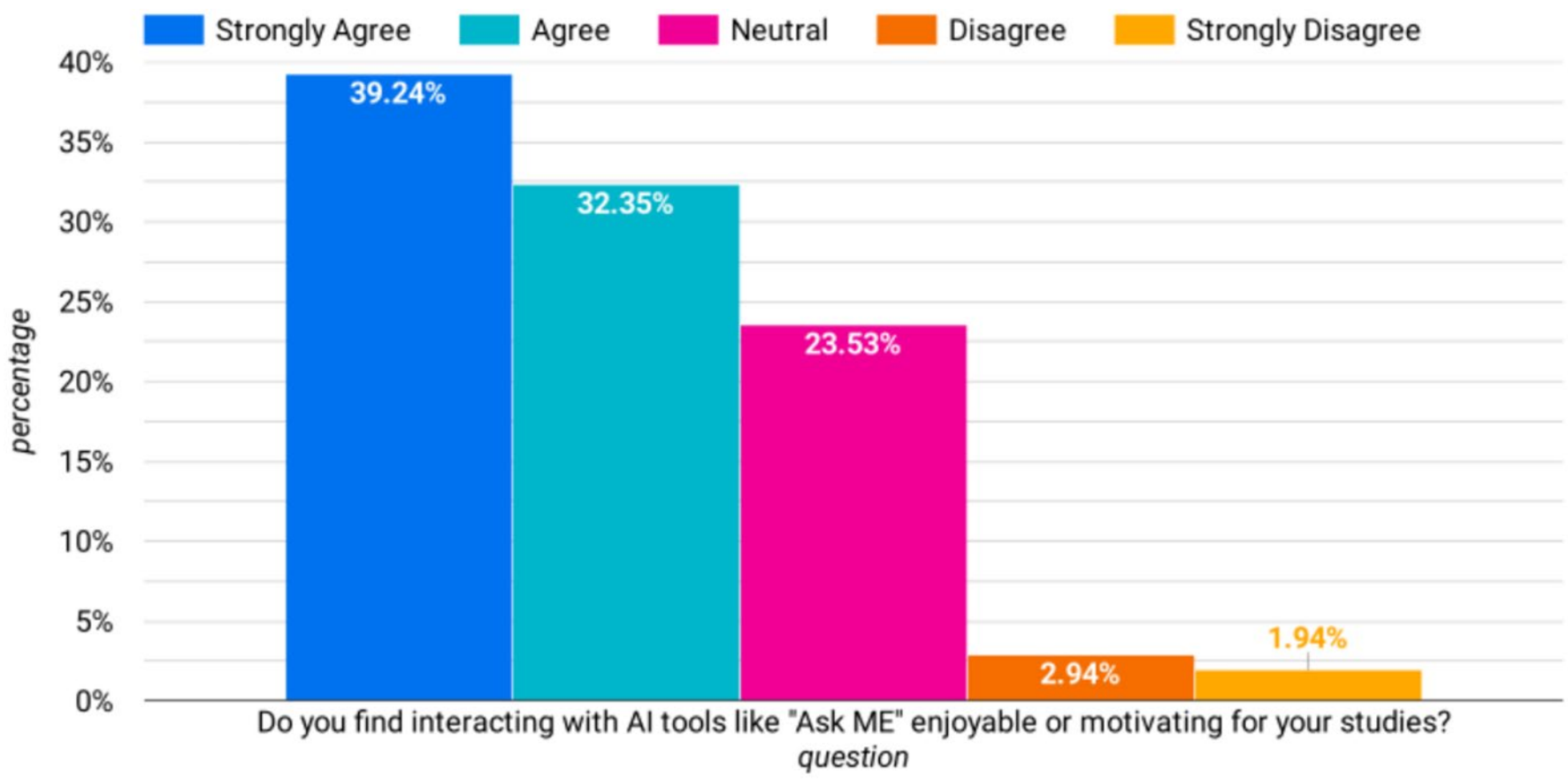

**Fig. 15** Student responses on enjoyment and motivation in using AI tools like Ask ME

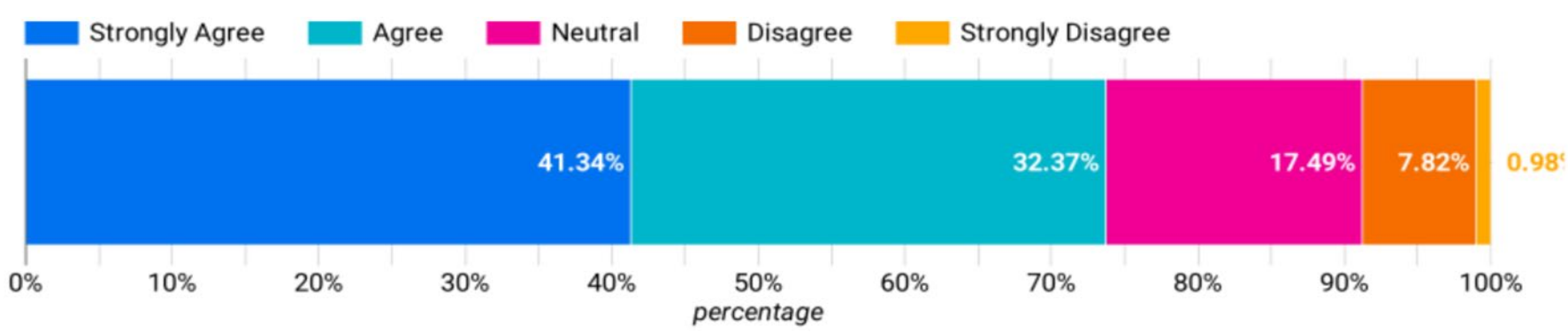

**Fig. 16** Student responses on the impact of AI tools like Ask ME in reducing classroom hesitation

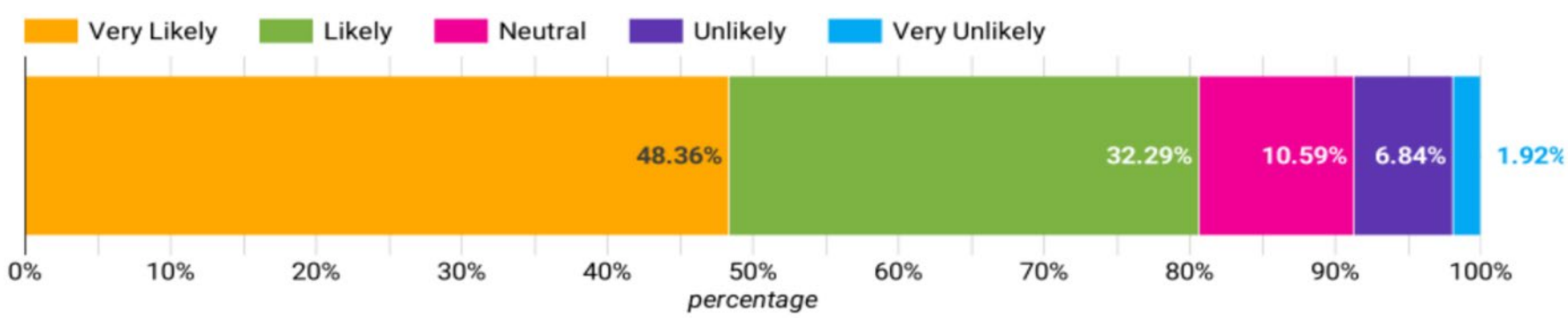

**Fig. 17** Student willingness to recommend AI tools like Ask ME for other courses

In terms of perceived educational benefits, a combined 75.51% of students rated AI-based tools in education as "*Very Important*" (49.04%) or "*Important*" (26.47%). More specifically regarding Ask ME, over 70% of respondents reported that the assistant encouraged them to explore more topics and questions than they otherwise would have (Fig. 19). A significant majority 77.57% believed that AI tools like Ask ME promote independent and self-directed learning (Fig. 20).

Furthermore, the integration of Ask ME with course content, facilitated by DCCI, was perceived as effective in improving understanding and engagement by 78.57% of students (Fig. 21) (H2). A total of 68.73% supported expanding AI to feedback/assessment tasks (Fig. 22), with 21.47% neutral and 9.8% dissenting. Qualitative feedback emphasized efficiency ("*faster response time*"), utility for missed content ("*i missed 2 lab classes so ask me help me cover those topics and also it helps me understand the course grading, tms user guidiance was helpful*"), and ease of access to content "*did not have to repeatedly look for a specific topic in slides; I could just ask Ask ME*" further validating H1's competence enhancement and H2's perceived usefulness.

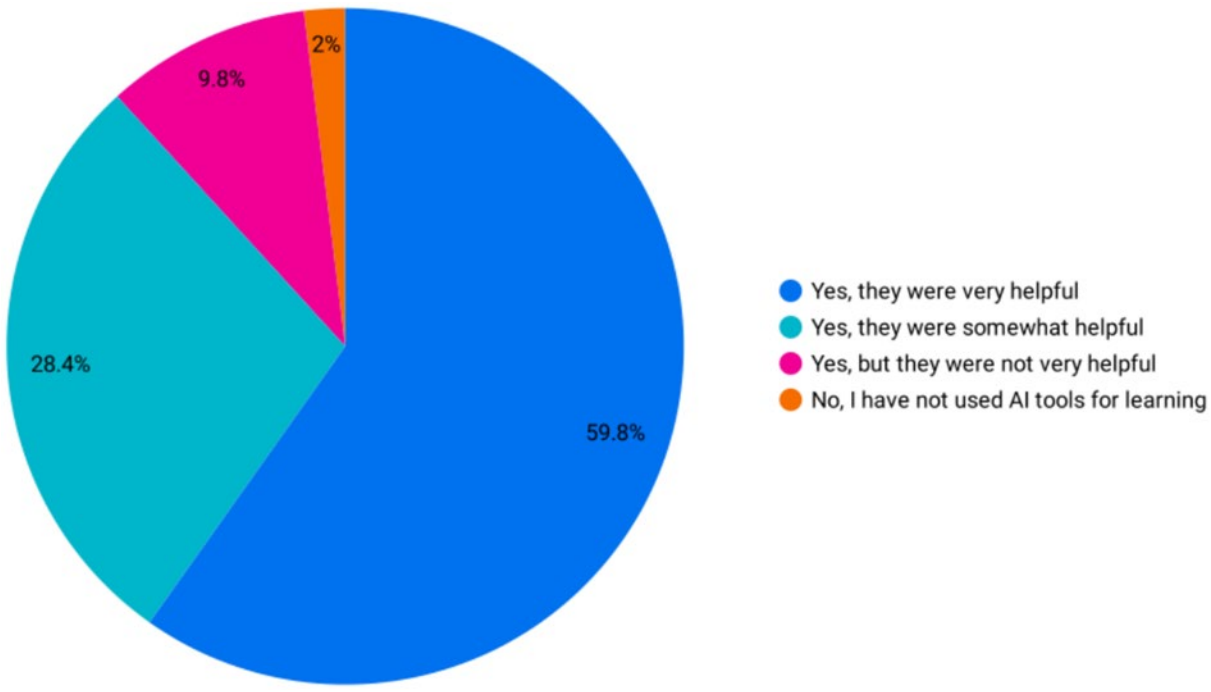

**Fig. 18** Student responses on the usefulness of other AI tools for learning

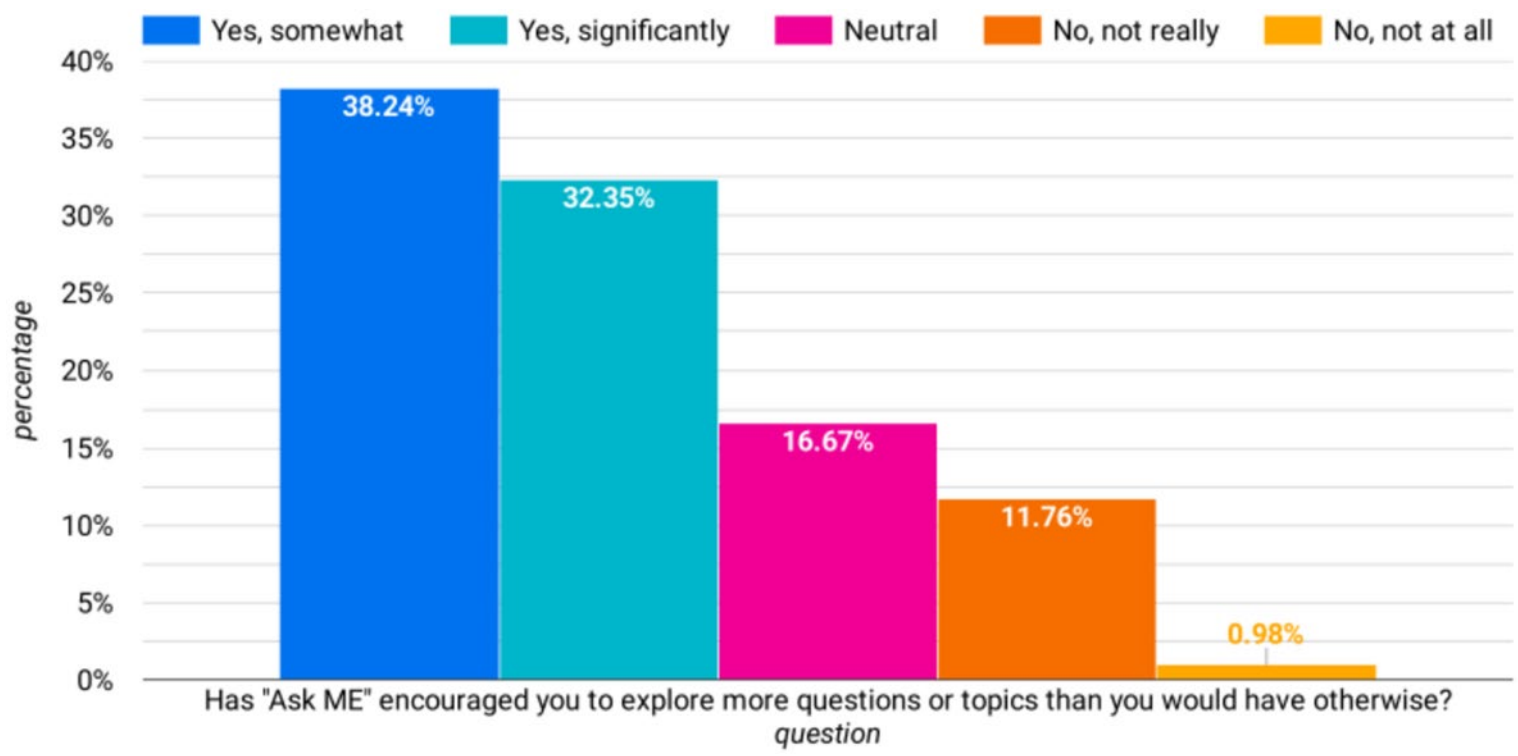

**Fig. 19** Student responses on AI's ability to encourage exploration and independent learning

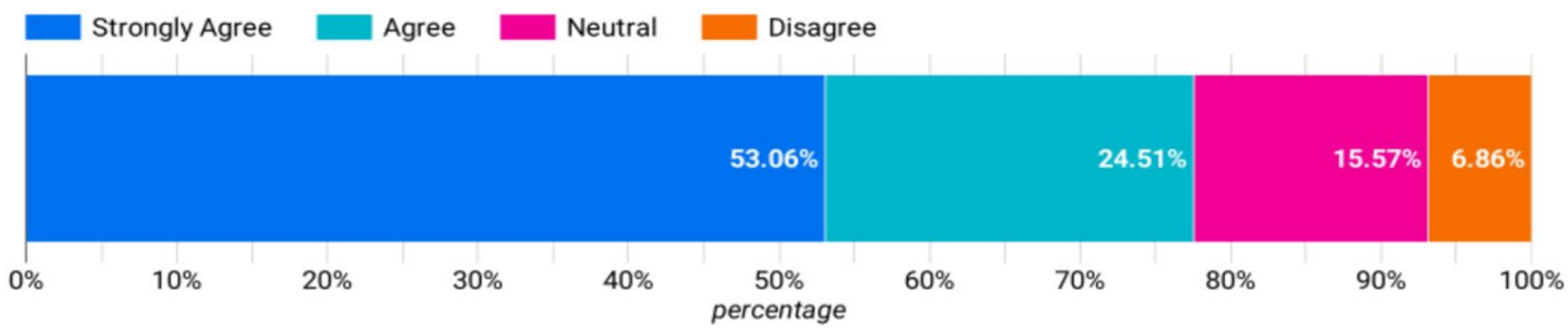

**Fig. 20** Student responses on AI tools promoting independent and self-directed learning

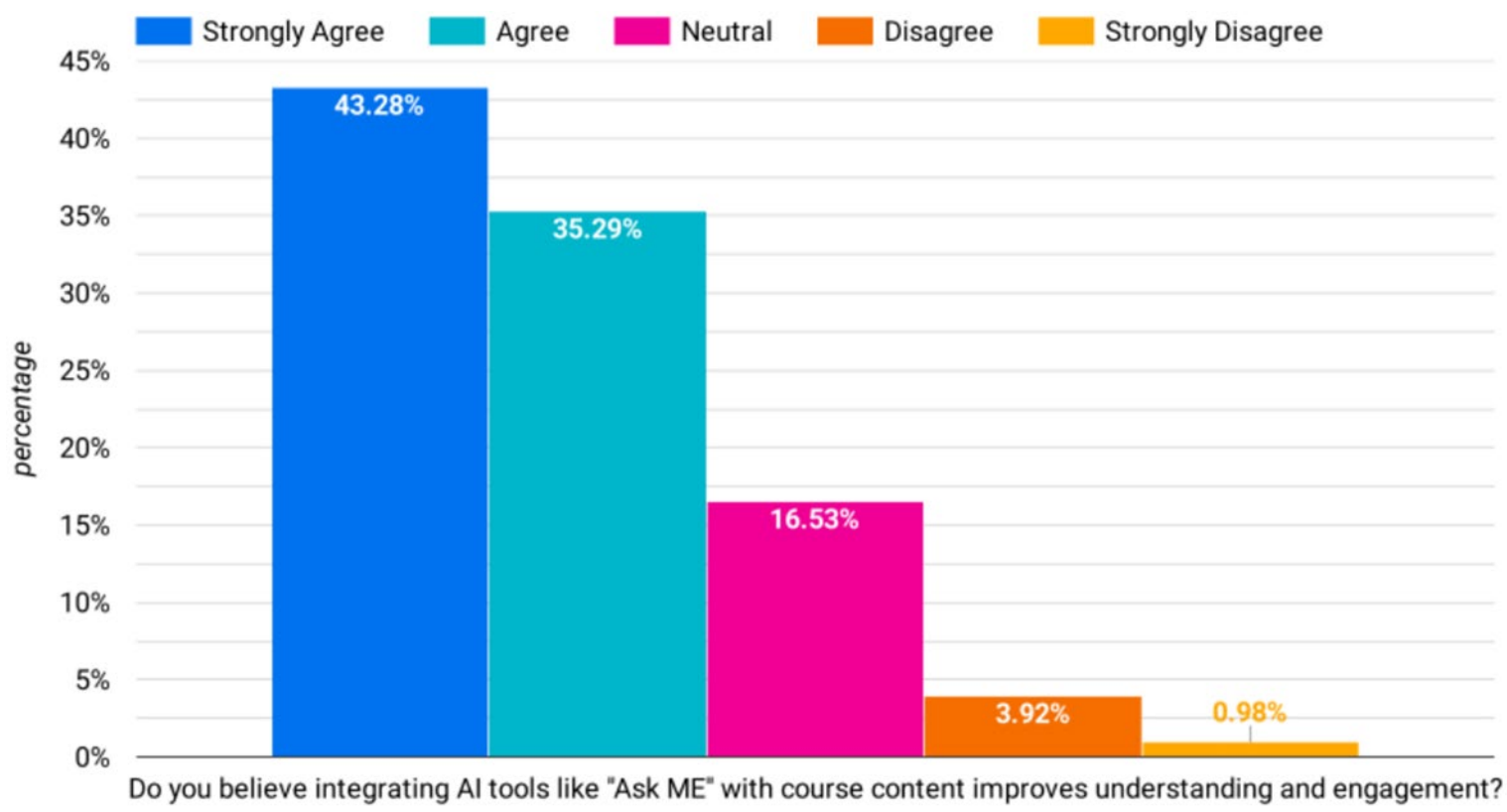

**Fig. 21** Student responses on the impact of AI tools like Ask ME on understanding and engagement

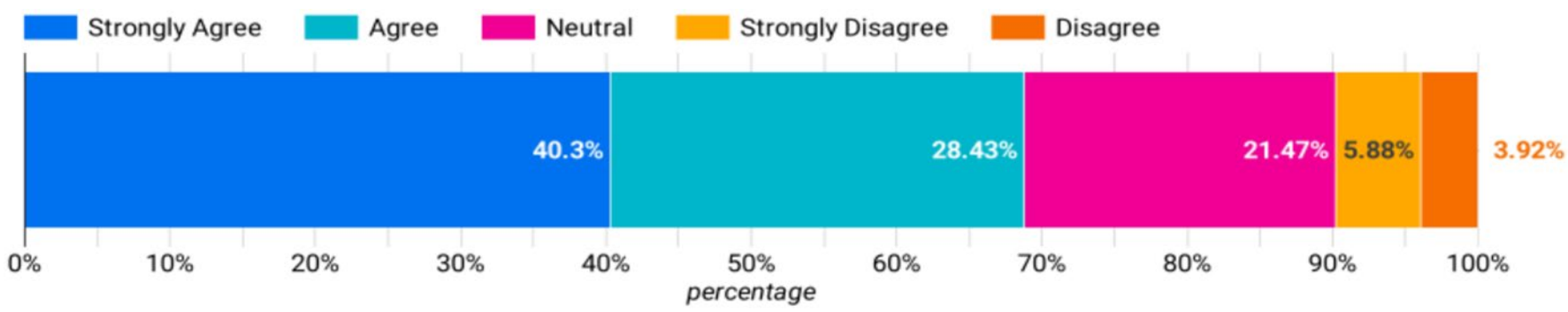

**Fig. 22** Student responses on expanding AI tools to feedback, assessment, and content generation

Preferences for asking questions to AI tools like *Ask ME* over directly approaching instructors were mixed: 10.78% of students chose "Always," 31.37% "Often," 38.24% "Sometimes," and 19.6% selected either "Rarely" or "Never".

## RQ3: Perceptions of AI risks and challenges

Despite the positive impacts, the survey responses regarding the potential risks and challenges of relying on AI tools in education reveal several concerns.

A predominant issue raised by respondents is over-reliance on AI, which they argue could diminish critical thinking, problem-solving abilities, and creativity. One participant warned that "*it may limit the thinking of students,*" while another remarked that "*these tools can take away the opportunity to think deeply from humans.*" Another observation noted that "*AI makes students lazier and lazier to search on the web or ask and understand questions from professors or mates.*"

Another significant concern is the accuracy of AI-generated information. Respondents noted that AI tools are susceptible to errors, outdated content, and incomplete responses, which may mislead students and disrupt the learning process. One participant emphasized that "*AI tools are not always correct and are bounded by the data they are trained on, so we cannot always rely on them,*" while another cautioned that "*malfunctioning in response may be a potential risk.*"

Ethical concerns, particularly plagiarism and academic dishonesty, were highlighted as potential risks. Students expressed apprehension that AI tools might be misused to bypass the learning process. For instance, one student commented, "*If not used well, the student could just copy and paste the solutions instead of learning from the to*ol".

Finally, the potential weakening of student–teacher interactions was raised as a concern. Respondents worried that students might increasingly depend on AI for answers instead of engaging directly with educators. One participant observed, "*instead of asking and interacting with the teacher, students might just completely depend on AI,*" potentially leading to a "*weaker student–teacher relationship*". These articulated concerns indicate an awareness of the complexities involved in integrating AI into educational environments.

## Discussion

The DCCI mechanism significantly enhanced the quality of LLM-generated responses by anchoring outputs to verified course content retrieved from Canvas LMS. This approach reduced hallucination risks, as evidenced by 90.4% of students perceiving responses as consistent with course materials (Fig. 10) and high satisfaction ratings for DCCI-dependent knowledge bases (like "Weekly Topics" (4.768/5) and "General Course Information" (4.641/5, Table 2). Perceived usefulness was further evidenced by 80.65% of students expressing willingness to recommend Ask ME (Fig. 17).

This underscores the effectiveness of the DCCI mechanism. By integrating course content from Canvas LMS with an LLM, DCCI enables the delivery of accurate, relevant, and timely human-like responses. These results align with prior research on LLMs' role in enhancing administrative efficiency and improving access to course-related information (Bonner et al., 2023; Tripathi et al., 2024).

These findings directly support H2, which posited that perceived usefulness (response accuracy) would mediate the relationship between DCCI's technical performance and student satisfaction. The DCCI mechanism's technical performance, ensuring that the bot retrieves verified content from structured course elements on Canvas LMS, significantly reduces the risk of hallucination a primary challenge with LLMs. This was a critical strength of Ask ME. By structuring retrieved content within the context window using prompt engineering, the mechanism ensures responses remain aligned with course materials, improving accuracy and reliability. Previous research has highlighted that hallucination in LLMs often occurs when models rely solely on their pre-trained knowledge without anchoring responses in authoritative sources (Z. Lin et al., 2024a, 2024b; Romera-Paredes et al., 2024). The implemented lightweight design of the context-aware retrieval approach in DCCI effectively constrains the LLM's outputs to dynamically retrieved content. The consistently high satisfaction ratings for the "Weekly Topics" and "General Course Information" knowledge bases (Table 2) validate that this technical performance translated into perceived usefulness (accurate, relevant responses), which in turn fostered high student satisfaction.

A key advantage observed is Ask ME's ability to streamline administrative support, effectively handling inquiries regarding course objectives, curriculum, schedules, grading policies, institutional procedures, and the TMS user manual, thereby potentially reducing the workload on educators and staff. This aligns with research demonstrating AI's capacity to automate responses to frequently asked questions, allowing educators and administrators to focus on complex pedagogical tasks (Adel et al., 2025; Ahmad et al., 2022; Sajja et al., 2023). Beyond administrative aid, Ask ME effectively assisted with course-related inquiries, including concept explanations, assignment guidance and clarifications on weekly topics and programming exercises, with the "Weekly Topic" knowledge base being the most frequently accessed, emphasizing its role in academic engagement. This aligns with studies on AI tutoring systems enhancing learning through personalized, on-demand support (C.-C. Lin et al., 2023; Mannekote et al., 2024).

H3 proposed that seamless LMS integration would reduce platform-switching, improving usability and engagement. The integration of Ask ME within Canvas LMS was a significant factor in enhancing accessibility and engagement. A majority of students (76.61%) found the integration helpful, and 78.06% agreed it reduced platform switching, minimizing cognitive and logistical barriers (Figs. 13, 14). This aligns with research on user-centered system integration, which emphasizes embedding support tools within existing learning environments to optimize usability and minimize disruption (Bennett et al., 2016; Conde-González et al., 2014) and findings by Alotaibi (2024) and Eltahir and Babiker (2024) on AI-LMS integration enhancing engagement, personalize learning, and learning outcomes in higher education. The substantial engagement metrics (14,746 queries, average session 25.32 min) further attest to the improved usability and engagement stemming from this integration. However, the presence of neutral and negative responses about the LMS integration may reflect individual preferences and highlights areas for improvement. For instance, qualitative feedback revealed some technical limitations, such as difficulties in locating the tool or preferences for a larger interface. These insights suggest opportunities to refine the user interface and further tailor the tool's functionality to diverse needs.

H1 posited that autonomy support and competence enhancement would positively correlate with student satisfaction and reduced hesitation. The 24/7 availability of Ask ME provided autonomy support, with students valuing the ability to get instant answers. This, coupled with *77.57%* of students believing AI tools promote independent learning (Fig. 20), supports the autonomy aspect of H1. Competence enhancement was evident from the high satisfaction with DCCI's context-aware responses (Table 2) and qualitative feedback like the tool "*explained me the tasks very good and make me happy to understand it without asking anyone*". This enhanced competence and autonomy correlated positively with overall student satisfaction (4.652/5).

Furthermore, the data strongly suggest that AI tools contribute to reducing classroom hesitation and fostering self-directed learning, key components of H1. Over 73.71% of students agreed that AI tools mitigate reluctance to ask questions in class (Fig. 16), supporting research on AI's role in creating inclusive learning environments (M. P.-C. Lin et al., 2024a, 2024b; Toyokawa et al., 2023). Studies have consistently shown that students often hesitate to seek help due to fear of judgment (Dyrbye et al., 2015; Horsch, 2006). AI-based assistants, such as Ask ME, have the potential to mitigate this concern by providing a non-judgmental platform where students can freely ask course-related questions without apprehension (Sajja et al., 2023). This was particularly evident in the emotional impact reported by students. Comments such as "*reduced anxiety from immediate assistance on simple questions*" highlight the tool's role in creating a supportive, judgment-free environment. This aligns with Farrelly and Baker's (2023) assertion that AI assistants can reduce hesitation by alleviating concerns about judgment. The observed increase in confidence is also supported by research indicating that immediate, autonomous access to learning resources enhances self-efficacy (Bandura, 1997; Lemos et al., 2017). The tool's ability to improve comprehension and efficiency also echoes findings by Sajja et al. (2024) and Gligorea et al. (2023). A key factor in this dynamic may be the support AI provides to students whose first language is not English, as they often face language-related challenges when seeking clarification in traditional English-language classroom settings. Research indicates that language proficiency significantly influences student participation, with non-native speakers frequently experiencing anxiety or hesitation due to concerns about grammar, pronunciation, or clarity (Coryell & Clark, 2009; Jahns, 2024). By offering a low-pressure environment for formulating questions and receiving structured explanations, AI tools can help bridge this gap and promote more confident engagement.

The third research question (RQ3) explored student perceptions of the risks and challenges associated with AI-driven tools. Survey responses revealed strong student endorsement of AI tools in education, with 75.51% rating them as "Very Important" or "Important," reflecting a growing acceptance of AI-driven assistance. This supports prior research emphasizing AI's role in providing personalized and adaptive support (Gligorea et al., 2023; Tapalova & Zhiyenbayeva, 2022) and fostering autonomous learning (Chang et al., 2023; Yim & Su, 2025).

However, this positive view was balanced by significant concerns. One major issue is over-reliance on AI. Respondents expressed worries that excessive dependence may diminish students' ability to engage in critical thinking and independent problem-solving. This aligns with scholarly arguments (Cui & Alias, 2024; Marrone et al., 2024) highlighting that excessive dependence on AI in education can hinder the development of deep thinking and creativity. Empirical research further supports this concern, with Gerlich (2025) identifying a negative correlation between frequent AI use and critical thinking skills, mediated by cognitive offloading. Collectively, this evidence suggests that an over-dependence on AI could undermine students' ability to develop essential cognitive skills.

Concerns regarding the accuracy of AI-generated information were also prominent, aligning with literature on LLM limitations (Ferrara, 2023). Students noted that AI tools can provide outdated or erroneous content, emphasizing the need for human oversight, reinforcing research advocating AI as a complement rather than a replacement (Ifenthaler et al., 2024; Renz & Vladova, 2021). Some users of Ask ME specifically expressed frustration over incorrect or incomplete answers, with comments such as "*wrong or incomplete answers can demotivate learning*" and "*way too complex answers that make me feel like I understand the subject less.*" These instances, though representing a smaller subset, were noted to potentially demotivate learning and may contribute to the 22.55% and 23.53% of respondents who selected "Neutral" for questions on anxiety reduction and motivation, respectively (Figs. 14, 15). For example, one respondent noted that the tool is helpful but "*cannot deal with all programming patterns well.*" These challenges stem in part from the fine-tuning process of the underlying language model used by Ask ME's fifth knowledge base, which is beyond the scope of this study. The fine-tuning was specifically targeted at program specifications and two primary programming patterns, summation and counting as an initial step. Expanding the training dataset to encompass a broader range of programming patterns may help address these limitations and enhance user satisfaction.

Ethical considerations further highlight the complexities of AI adoption in education. Concerns about plagiarism and academic dishonesty were prominent, with students cautioning that AI tools could be misused to bypass learning rather than enhance it, reinforcing findings that such tools may encourage uncritical reproduction of knowledge rather than meaningful engagement (Cotton et al., 2023; Gruenhagen et al., 2024).

Finally, apprehensions regarding weakened student–teacher interactions reflect research suggesting that while AI enhances accessibility, it should not replace the formative role of educators (Arango et al., 2024; Cope et al., 2020). Variations in student preferences regarding AI versus instructor interaction (with only 10.78% "Always" preferring AI) suggest that while AI tools are valued, they are not perceived as full replacements for human educators, aligning with research advocating for human-AI collaboration, where AI supplements rather than supplants traditional instructional methods (Atchley et al., 2024; Ifenthaler & Schumacher, 2023).

Thus, while AI-powered tools like Ask ME offer valuable support, their effectiveness depends on strategic integration, ethical safeguards, and a pedagogical framework that prioritizes human-AI collaboration over substitution. The occasional inquiries related to mental health and social issues (e.g., "*I'm just lonely. Can you talk to me?*"), though not a primary function, also hint at a broader, perhaps unmet, need for accessible support, which AI might partially address, but which also underscores the irreplaceable nature of human connection and professional support in these areas.

## Conclusion, limitation, future work

### Conclusion

This study examined the effectiveness of the DCCI mechanism in enhancing generative AI-driven educational assistance through the Ask ME assistant bot within Canvas LMS. Empirical results from interaction logs and survey responses robustly address the research questions and hypotheses, demonstrating DCCI's capacity to improve LLM response quality, student satisfaction, and engagement by addressing both administrative and course-related inquiries efficiently.

The high user satisfaction rating (4.652/5), along with 90.4% of students perceiving the responses as consistent with course materials, underscores Ask ME's success in delivering accurate, context-aware, human-like responses, thereby validating H2. By anchoring outputs to dynamically retrieved course content, DCCI minimized hallucination risks a critical advancement over traditional LLM reliance on pre-trained knowledge alone.

H1 and H3 were similarly validated. Autonomy support (24/7 access) and competence enhancement (context-aware responses) correlated with reduced hesitation (73.71% agreement) and increased self-directed learning (77.57%), reinforcing SDT. Seamless Canvas LMS integration reduced platform-switching (78.06%), enhancing usability and engagement (14,746 queries, 25.32-min sessions), in line with TAM principles.

However, while the findings demonstrate the potential benefits of AI integration in education, they also reveal important considerations. Over-reliance on AI, concerns about accuracy, ethical risks (such as plagiarism), and the potential weakening of student–teacher interactions emerged as notable challenges voiced by students. These concerns emphasize the need for balanced AI adoption strategies, ensuring that AI tools complement rather than replace traditional instructional methods. Effective human-AI collaboration, coupled with institutional safeguards and ethical guidelines, is essential to maximizing AI's benefits while mitigating its risks.

These findings contribute to the broader discourse on AI-enhanced education, demonstrating the importance of LLM-driven course content integration with LMSs, responsible AI deployment, and ethical oversight in shaping the future of AI-assisted learning environments.

## Limitations

While this study demonstrates the effectiveness of the Ask ME assistant bot and the DCCI mechanism in delivering context-aware, accurate, and timely responses on Canvas LMS, several limitations must be acknowledged.

First, its implementation was limited to a single course, restricting generalizability. Expanding deployment across diverse academic settings would provide deeper insights into its adaptability and scalability.

Second, DCCI's reliance on the LLM's context window constrains the amount of information processed per interaction. While Gemini 1.5 Flash's 2-million-token capacity mitigates this issue and effectively accommodates extensive course materials, models with smaller context windows require careful content selection to prevent truncation and response inaccuracies.

Third, Ask ME currently supports only text-based interactions, limiting its applicability in disciplines requiring visual elements such as diagrams or flowcharts. Integrating multimodal capabilities would enhance its functionality.

Fourth, data privacy remains a concern, as Gemini 1.5 Flash processes course content in an external cloud environment, raising security and compliance issues, particularly when handling sensitive academic materials that course owners may be reluctant to share externally. While hosting open-source LLMs locally could enhance data control, this requires significant computational resources, posing challenges for institutions with limited infrastructure.

Finally, the study relies on self-reported data. While user satisfaction ratings, survey responses, and interaction logs provide valuable insights, they may not fully capture long-term learning outcomes, behavioral changes, or cognitive engagemt. Future research should incorporate longitudinal studies, qualitative interviews, and performance-based assessments.

### Recommendations for enhancing Ask ME

Survey responses reveal a generally positive sentiment toward “Ask ME,” with participants frequently describing it as “*helpful*” and “*useful*.” However, several areas for enhancement were identified. Many respondents emphasized the need for expanded data integration, broader knowledge bases, and the option to upload files. Customization and specificity of responses also emerged as critical areas for improvement. For example, a request to “*add the ability to generate mock tests with solutions*”.

Participants also emphasized the importance of enhancing the tool’s interactive and problem-solving capabilities. Suggested improvements include features such as “edit” and “stop processing” buttons, as well as support for multimedia inputs like diagrams and structograms. A user also highlighted the need for more adaptable responses, noting that “*It would be helpful if it could solve problems with detailed explanations,*”

Another recurring theme was the desire for expansion to other courses. Participants indicated interest in broader adoption, with one suggesting, “*It would be great if other classes also had the ’Ask ME’ tool,*” and another recommending, “*It should be used across the entire faculty.*” Increasing awareness of the tool’s capabilities was also noted, with one user urging, “*Encourage students more to use it, as many are unaware of its potential.*”

### Future works

Based on student feedback and observed usage patterns, several areas for future development have been identified to enhance Ask ME’s integration, functionality, and user experience.

A key recommendation is the expansion of Ask ME beyond a single course to function as a faculty-wide AI assistance service. Implementing a scalable mechanism at the application level within the faculty’s Canvas LMS environment would enable seamless accessibility across all courses. The mechansim will enable Ask ME for each course to dynamically retrieve course-specific content, acting as an intelligent assistant for both course administrators and curriculum-related inquiries, creating a consistent, adaptive, and personalized learning assistant across disciplines. This mechanism could be structured as a framework or standard for incorporating LLM technology into course content within LMS platforms, with broader features and functionalities. Additionally, broadening Ask ME’s knowledge base to include user manuals for tools used in courses, such as specification writing tools that use specialized syntax and notation, would address specific student challenges. Interaction logs indicate that students often struggle with properly using these tools, suggesting that incorporating user manuals and guidance into Ask ME could enhance their learning experience and efficiency.

Beyond responsive assistance, the integration of LLM with the DCCI mechanism holds potential for automating educational tasks such as content generation, assessment design, and personalized learning materials. Future work can explore its application in generating assignments, quizzes, and instructional content that align with course contnet and curriculum topics dynamically retrieved from LMS. Additionally, leveraging student profiles to personalize learning pathways would allow for tailored content recommendations that align with individual academic progress and needs.

Another critical advancement involves enhancing Ask ME's ability to process multimodal content. Currently limited to text-based queries. Future developments can focus on integrating diagrams, flowcharts, structograms, and images, enabling Ask ME to interpret, generate, and incorporate visual elements into responses. Also Integrating multimodal AI processing will further align Ask ME with advanced AI tutoring systems, enhancing its ability to personalize responses and adapt to diverse learning needs.

Improving Ask ME's user interface and interactive features is another area of focus to enhance accessibility and engagement. Future enhancements will aim to introduce customizable interaction modes that allow users to choose between different response formats, such as concise versus detailed explanations, and implement enhanced conversational memory to retain contextual awareness over longer interactions. Additionally, Incorporating contextually relevant follow-up questions into responses will encourage students to explore related topics, deepen their understanding, and enhance their engagement with course content.

As AI continues to play an increasing role in educational environments, it is imperative to ensure ethical, transparent, and pedagogically sound implementation. Future research should focus on developing institutional guidelines, refining oversight mechanisms to address concerns related to plagiarism, over-reliance on AI, and exploring adaptive AI behavior that recommends human-instructor involvement when necessary.

**Author contributions** Conceptualization, KM and MT-S; methodology, KM; software, KM; validation, KM; formal analysis, KM; investigation, KM; resources, KM and MT-S; data curation, KM; writing—original draft preparation, KM; writing—review and editing, KM and MT-S; visualization, KM; supervision, MT-S, KM; project administration, KM and MT-S All authors have read and agreed to the published version of the manuscript.

**Funding** Open access funding provided by Eötvös Loránd University. No funding was received to assist with the preparation of this manuscript. The authors have no relevant financial or non-financial interests to disclose.

**Data availability** The data presented in this article are available within the text. Additional data can be requested from the corresponding author.

## Declarations

**Conflict of interests** The authors declare no competing interests.

**Ethics approval** The study was conducted in accordance with the Declaration of Helsinki, and approved by the Ethics Committee of ELTE Faculty of Informatics (Approval Code: IK/164/1(2024)). Approval Date: 26 August 2024.

**Informed consent statement** Informed consent was obtained from all subjects involved in the study.

**Kovan Mzwri** is a Ph.D. candidate in Technology-Enhanced Education at Eötvös Loránd University, where he also serves as an assistant lecturer. His research focuses on integrating Generative AI (GAI) and large language models (LLMs) into learning environments and learning management systems (LMSs) to enhance autonomous learning, facilitate personalized learning experiences, and provide student and instructional support. He is particularly interested in developing innovative pedagogical agents, with a focus on chatbots and conversational agents. He earned his Master's degree in Computer Science from Eötvös Loránd University in 2017 under the Stipendium Hungaricum scholarship program. His expertise spans GAI and LLM applications in education, software development, AI-driven learning solutions, LMS integration, web development, and university teaching. His current research explores the role of GAI and LLMs in education, the impact of chatbots in technology-enhanced learning, and AI-driven student mentoring to improve student support, engagement, and personalized learning experiences. Kovan Mzwri has been awarded the Stipendium Hungaricum scholarship for his Ph.D. studies. His work has been presented at academic conferences and published in high-quality peer-reviewed journals.

**Dr. Márta Turcsányi-Szabó** Ph.D. is an associate professor, head of STEAM Lab at Eötvös Loránd University. Her research area is Technology Enhanced Learning, teaching several courses (taught at UniVie, Vienna and HBMSU, Dubai as well), participating and leading R&D projects on local, EU and UNESCO in the area. She received several prizes (Rezso Tarján prize for popularizing Informatics, iELA Gold prize in Mobile category, Tempus STEM award for best practice in Higher Education, Multimedia Ring for Teacher Education and Szent-Györgyi Albert state prize for establishing school of thought). She is presently the dean's commissioner for Educational Innovation at the Faculty of Informatics.

## Authors and Affiliations

### Kovan Mzwri[1] · Márta Turcsányi-Szabo[2]

✉ Kovan Mzwri
kovan@inf.elte.hu

Márta Turcsányi-Szabo
tszmarta@inf.elte.hu

1 Doctoral School of Informatics, Eötvös Loránd University, Pázmány Péter Stny. 1/C, Budapest 1117, Hungary

2 Department of Media and Educational Technology, Faculty of Informatics, Eötvös Loránd University, Pázmány Péter Stny. 1/C, Budapest 1117, Hungary